\documentclass[10pt,twoside]{article}
\usepackage{txfonts}\usepackage[full]{textcomp}
\usepackage[numbers]{natbib}
\newcommand{\binom}[2]{{#1 \choose #2}}
\newcommand{\xnum}[2]{#1$\,\cdot\,$$10^{#2}$}
\newcommand{\xpten}[1]{$10^{#1}$}

\newenvironment{lquote}{\begin{list}{}{}\item[]}{\end{list}}
\begin{document}

\thispagestyle{empty}

\begin{center}
 {\Large A numerical analysis of Quicksort: \\ How many cases are bad
  cases?} \\[0.4cm]

\renewcommand{\thefootnote}{\fnsymbol{footnote}}

{\large Guido Hartmann} \\[0.24ex]
 {\em Kiel University of Applied Sciences,
 Germany\footnote{Fachhochschule Kiel -- University of Applied Sciences,
 Fachbereich Informatik und Elektrotechnik, Grenzstra{\ss}e 5,
 \\ 24149 Kiel, Germany. \ E-mail: guido.hartmann\symbol{64}fh-kiel.de}
 } \\[0.26ex]

July 15, 2015
\end{center}

\medskip

\begin{center}
 ABSTRACT
\end{center}
{\small
We present numerical results for the probability of bad cases for
Quicksort, i.\,e. cases of input data for which the sorting cost
considerably exceeds that of the average. Dynamic programming was used
to compute solutions of the recurrence for the frequency distributions
of comparisons. From these solutions, probabilities of numbers of
comparisons above certain thresholds relative to the average were
extracted. Computations were done for array sizes up to \textit{n} = 500
elements and for several methods to select the partitioning element,
from a simple random selection to what we call ``recursive median of
three medians.'' We found that the probability strongly depends on the
selection method: for \textit{n} = 500 and a threshold 25\% above
the average number of comparisons it ranges from \xnum{2.2}{-3} to
\xnum{3.0}{-23}. A version of Quicksort based on the recursive median of
medians approach is proposed, for which our data suggest a worst case
time complexity of $O(n^{1.37})$.

\medskip

\noindent
 KEYWORDS: Quicksort; numerical analysis; bad cases; probability;
 frequency distribution of comparisons; \linebreak recursive median of
 three medians; implementation
}

\vspace{0.18in}

\section*{\textnormal{1. \ Introduction}}

\noindent
In 1961, C.\,A.\,R.\,Hoare published a sorting method which he called
``Quicksort" \cite{hoare61, hoare62}. The name is not an exaggeration:
a careful implementation of Quicksort performs, in the average case and
for large numbers of elements, better than any other of the established
comparison based array sorting algorithms. A further advantage is that
the sorting is done in place.

However, there remains the annoyance that the time complexity of
Quicksort is actually $O(n^2)$, and $O(n \log n)$ only on average.
For instance, sorting of one million elements may need a factor of
18,000 longer than expected if, unfortunately, a very ``bad case'' for
the algorithm is encountered.

For the practitioner who has to make judgement on the usability of a
Quicksort variant for a given application it would be desirable to know
probability values, at least orders of magnitude, for such cases. What
is, for a given number of elements, the relative number of cases needing,
say, more than 150\% of the average sorting time? How does such a ratio
depend on the number of elements, how on the threshold factor relative
to the average, how on the Quicksort version used?

As yet, theoretically derived formulas for the probability  of bad cases
(\cite{mcdia1, mcdia2, filljan}) hardly meet the expectations of the
practitioner. Either it is necessary to fit a parameter to simulation
results first, or only a bound is given that may be orders of magnitude
above realistic values.

In this article an attempt is made to give answers to those questions,
based only on numerical computations. We present results for the
probability of cases for which the cost will be above certain thresholds
relative to the average. As criterion for the cost we confine ourselves
to the number of comparisons of elements, the basic operation that
determines the time complexity of Quicksort. In addition, we restrict
ourselves to arrays of pairwise different elements. However, we attach
importance to base our computations on algorithms which are practically
usable, and in particular perform satisfactorily also if elements of
equal values are present.

\pagebreak

To get our results, we numerically computed the complete frequency
distributions of comparisons for up to 500 elements. We also studied the
influence of different methods for the selection of partitioning
elements on the probability of bad cases. As a by-product of our
investigations, we are able to propose a Quicksort version that has
extremely small probability values for bad cases and a time complexity
below $O(n^2)$.

Frequency distributions of comparisons have already been computed by
Eddy and Schervish \cite{edsch} for numbers of elements up to 132, and
by McDiarmid and Hayward \cite{mcdia2} for 100 elements; in both cases,
however, without extracting the probabilities we are interested in.

\medskip

The remainder of this paper is organized as follows.
We give a short recollection of the Quicksort algorithm and show the
partitioning method on which our computations are based in Section 2.
Following in Section 3, we present the fundamental formulas, particularly
for the determination of frequency distributions. After this, in Section
4 we discuss the models and approximations we used, including the
recursive median of three medians approach. In Section~5, some technical
details of the computations are given. The results are presented in
Section~6. Following in Section 7, characteristics of several methods
for partitioning are discussed. In Section 8 we comment on our Quicksort
implementation. Finally, Section 9 concludes the paper.
A listing of our proposed Quicksort version is given in the appendix.

\bigskip

{\noindent \fboxsep0.2cm
\fbox{\parbox{13.90cm}{
Note that throughout this article variable name \textit{n} will
exclusively be used for the number of array elements to be sorted.
}}}

\medskip

\section*{\textnormal{2. \ Quicksort fundamentals}}

\subsection*{\textnormal{\textsl{2.1 \ The basic algorithm}}}

Quicksort is a recursive algorithm following the divide-and-conquer
paradigm. It may be formulated as follows:

 \begin{lquote}

 {\sl Recursion basis:}

 If \textit{n} $\le$ 1, nothing has to be done.

 \medskip

{\sl Recursive part (if \textit{n} $>$ \textup{1}):}

 Choose one of the array elements as ``partitioning element'' p.
 Partition the array, i.\,e. rearrange its elements so that an order
\begin{equation}
 \underbrace{\mbox{elements}\le \mbox{p,}}_{\mbox{subarray 1}} \
 \mbox{one element = p,} \ \underbrace{\mbox{elements} \ge
 \mbox{p}}_{\mbox{subarray 2}}
\end{equation}
 or
\begin{equation}
 \underbrace{\mbox{elements}\le \mbox{p,}}_{\mbox{subarray 1}} \ \
 \underbrace{\mbox{elements} \ge
 \mbox{p}}_{\mbox{subarray 2}}
\end{equation}
 is achieved. Only in the case of arrangement (1) one of the subarrays
 may be empty. Sort each of the two subarrays by recursion.

\end{lquote}

\begin{figure}
\vspace*{0.4cm}
{\small
\begin{verbatim}
   /* ipart: index of the partitioning element */
      partval = a[ipart];
      i = 0;
      j = n-1;
      a[ipart] = a[j]; a[j--] = partval;
      done = FALSE;
      do {
         while (a[i] < partval)
            i++;
         while (i < j && partval < a[j])
            j--;
         if (j <= i)
            done = TRUE;
         else {
            temp = a[i]; a[i++] = a[j]; a[j--] = temp;
         }
      } while (!done);
      a[n-1] = a[i]; a[i] = partval;

   /* subarray 1: a[0..(i-1)],  subarray 2: a[(i+1)..(n-1)] */
\end{verbatim}
}
\vspace{-8.13cm}
{\parindent0em \setlength{\unitlength}{1.0cm} \begin{picture}(14.31,8.2)
\put(0,0){\framebox(14.31,8.2){}}
\end{picture}
}
\begin{center} {\small Figure 1. Partitioning algorithm.
\verb+FALSE+ and \verb+TRUE+ are symbols for 0 and 1, respectively.}
\end{center}
\end{figure}

There are variants in which small subarrays are sorted by a simpler
method; in this case the recursion basis has to be extended. Two
details have been left out in the formulation above:
\begin{enumerate}
\item The method of choosing the partitioning element. \\[-1.6em]
\item The partitioning method.
\end{enumerate}

Our computations have been carried out for several methods of choosing
the partitioning element, these will be discussed in Section 4. Point 2
will be dealt with in the following subsection.

\medskip

\subsection*{\textnormal{\textsl{2.2 \ Algorithm for partitioning}}}

For our computations we assumed a partitioning method similar to that
given by Sedgewick \cite{sedg98}, program 7.2. Figure 1 shows our
algorithm, formulated in C. An explanation for the choice of this
version will be given in Section 7. The result is always arrangement (1)
as displayed in Section 2.1, where one of the subarrays may be empty.
The number of comparisons needed is either \textit{n}$-$1 or \textit{n}.
For distinct values of elements, it turns out that if the final index of
the partitioning element is $i$, the respective probabilities
$q^{(c)}_n(i)$ for the numbers of comparisons $c$ are
\begin{equation}
  q^{(n-1)}_n(i) \ = \ \frac{n-1-i}{n-1} \qquad \mbox{and} \qquad
  q^{(n)}_n(i) \ = \ \frac{i}{n-1}.
\end{equation}

\section*{\textnormal{3. \ Fundamental formulas}}

Our computations are based on the assumption that partitioning within
Quicksort is done by the algorithm shown in Figure 1. For pairwise
different elements, this will always lead to an arrangement
\begin{equation}
 \underbrace{\mbox{elements} < \mbox{p,}}_{\mbox{subarray 1}} \
 \mbox{p,} \ \underbrace{\mbox{elements} >
 \mbox{p}}_{\mbox{subarray 2}} .
\end{equation}

Throughout this and the following section, variable name $i$ will be
used exclusively for the index of the partitioning element p after the
partitioning process (indices numbered from 0).

\medskip

Let $f_n$ be the distribution of the frequencies of comparisons
needed by Quicksort to sort arrays of \textit{n} pairwise different
elements (\textit{n} $\ge$ 0). Precisely:
counting over all permutations, $f_n(j)$ is the number of cases for
which $j$ comparisons of elements are needed. So $f_n$ is a function
\begin{equation}
 f_n : \mathbb{N}_0 \longrightarrow \mathbb{N}_0
 \qquad\mbox{satisfying}\qquad
 \sum_{j \ge 0} f_n(j) \ = \ n!
\end{equation}
 ($\mathbb{N}_0$: the set of natural numbers including 0).
 The distributions may be computed from the recurrence

\begin{eqnarray}
 f_0(0) \!\! & = \!\! & f_1(0) \ = \ 1, \\
 f_0(j) \!\! & = \!\! & f_1(j) \ = \ 0  \qquad\mbox{for} \ \, j > 0, \\
 f_n \!\! & = \!\! & d_n \ast \frac{1}{2} (\delta_{n-1}+\delta_{n})
 \ast  \sum_{i=0}^{n-1} \binom{n-1}{i}\,p_n(i) \
  f_i \ast f_{n-1-i} \qquad\mbox{for} \ \, n > 1.   \label{rec1}
\end{eqnarray}

Equations (6) and (7) define the basis and express the fact that
cases \textit{n} = 0 and \textit{n} = 1 do not need any comparison.
In equation (\ref{rec1}) the recurrence for the distribution $f_n$ as
a whole is formulated. Here $p_n$ is a function such that $p_n(i)/n$
means the probability that the element chosen for partitioning will be
at index position $i$ after the rearrangement. We assume the symmetry
$p_n(i) = p_n(n$$-$$1$$-$$i)$ and require
\begin{equation}
 \sum_{i = 0}^{n-1} p_n(i) \ = \ n,
\end{equation}
 so that for the case of randomly chosen partitioning elements
 $p_n(i)$ is 1 for $0 \le i < n$ and thus could be left out.
 The binomial coefficient takes numbers of cases according to equation
 (5) into account.

Symbol $d_n$ denotes the distribution of the numbers of comparisons
that are needed only for the selection of a partitioning element. In
detail: $d_n(j)$ is the probability of needing $j$ comparisons of
elements to find the partitioning element. This function is normalized
as
\begin{equation}
 \sum_{j \ge 0} d_n(j)  \ = \ 1.
\end{equation}

 For any $k \in \mathbb{N}_0$, function $\delta_k$ is defined as
 \begin{equation}
 \delta_k : \mathbb{N}_0 \longrightarrow \mathbb{N}_0 \qquad
 \delta_k(j) \
 \stackrel{\mathrm{def}}{=} \
 \left\{ \begin{array}{l} 1 \quad\mbox{for}\quad j = k, \\ 0
 \quad\mbox{for}\quad j \not= k. \end{array} \right.
\end{equation}
 The result of a convolution of $\delta_k$ with another function will be
this function shifted by $k$. In our recurrence equation, expression
$(\delta_{n-1}+\delta_{n})/2$ addresses the comparisons needed for the
partitioning process itself. Symmetry has be used to derive this simpler
expression from the original
$(q^{(n-1)}_n(i)\,\delta_{n-1}+q^{(n)}_n(i)\,\delta_{n})$ within the sum
(cf. equations (3)).

Finally, symbol $\ast$ in the recurrence equation represents the
convolution operator, i.\,e. for functions
 $g, h : \mathbb{N}_0 \longrightarrow \mathbb{R}$
\begin{equation}
 (g \ast h)(j) \ \stackrel{\mathrm{def}}{=} \ \sum_{k=0}^j g(k)\,h(j-k)
 \qquad\mbox{for} \ \, j \ge 0.
\end{equation}

\medskip

 The average number (expected value) of comparisons needed by Quicksort
\begin{equation}
 \overline{C}_n \ \stackrel{\mathrm{def}}{=} \ \frac{1}{n!} \
 \sum_{j \ge 0} j f_n(j)
\end{equation}
 may be computed separately from the recurrence
\begin{equation}
 \overline{C}_n \ = \ \overline{C}^{\mbox{\scriptsize {\,select}}}_n
 + n - \frac{1}{2} +
 \frac{2}{n} \ \sum_{i=0}^{n-1} p_n(i)\:\overline{C}_i
   \qquad\mbox{for} \ \, n > 1,        \label{rec2}
\end{equation}
 where $\overline{C}^{\mbox{\scriptsize {\,select}}}_n$ is the average
 number of comparisons to select a partitioning element, i.\,e.
\begin{equation}
 \overline{C}^{\mbox{\scriptsize {\,select}}}_n
 \ \stackrel{\mathrm{def}}{=} \ \sum_{j \ge 0} j d_n(j).
\end{equation}

 For the maximum number of comparisons
\begin{equation}
 \widehat{C}_n \ \stackrel{\mathrm{def}}{=}
 \max_{j \ge 0 \wedge f_n(j) > 0} \, j
\end{equation}
 we have the recurrence
\begin{equation}
 \widehat{C}_n \ = \ \widehat{C}^{\mbox{\scriptsize {\,select}}}_n + n +
 \max_{0 \le i < n \wedge p_n(i) > 0} (\widehat{C}_i
  + \widehat{C}_{n-1-i}) \qquad\mbox{for} \ \, n > 1,     \label{rec3}
\end{equation}
where $\widehat{C}^{\mbox{\scriptsize {\,select}}}_n$ denotes the maximum
number of comparisons to select a partitioning element.

\section*{\textnormal{4. \ Models and approximations}}

In recurrence (\ref{rec1}), functions $p_n$ and $d_n$ have yet to be
fixed. We did our computations for several methods of selecting the
partitioning element; the two functions depend on the respective method.
Additionally, we analyzed an extended version of the Quicksort algorithm
where an insertion sort is used for small arrays. These different
versions and the respective approximations used in the computations will
be classified as five models, described in the following.

We are interested in the frequencies of bad cases, which result from
summations of large numbers of comparisons during traversal of the
recursion tree for recurrence (8). Any approximations we used in the
computations were chosen as such that rather some overestimation than
underestimation of these cases could occur.

\subsection*{\textnormal{\textsl{4.1 \ Model 1: Simple}}}

In the simplest version, the partitioning element for an array of
\textit{n} elements is always taken from a fixed position, usually at
index $\lfloor n/2 \rfloor$. Since we are counting over all
permutations of \textit{n} pairwise different elements, this is
equivalent to a random choice where each element has the same
probability of being selected. Hence we have
\begin{equation}
   p_n(i) \ = \ 1 \qquad\mbox{for} \ \, 0 \le i < n.
\end{equation}

No comparisons are needed for this simple selection, so the convolution
with $d_n$ can simply be left out.

\subsection*{\textnormal{\textsl{4.2 \ Model 2: Median of three}}}

Quicksort performs most efficiently if the two subarrays which result
from partitioning are equal (or almost equal) in size. Therefore
selection methods have been invented to increase the probability of
the partitioning element to end up at a position near the middle of
the array.

Hoare \cite{hoare62} suggested to take a small sample of array elements
and use its median for partitioning. In 1969, Singleton \cite{singl}
published a version that incorporates method ``median of three
elements'' (or ``median of three'' for short). From the first, middle
and last element in the array the median valued one is chosen. In
contrast to Singleton's version, we assume that the median is determined
without an explicit sort among the three elements. According to our
model, method ``simple'' is used for \textit{n} = 2. For \textit{n}
$\ge$ 3, median of three leads to
\begin{equation}
p_n(i) \ = \ a_n \ i (n-1-i) \qquad\mbox{for} \ \, 0 \le i < n,
\end{equation}
 where
\begin{equation}
a_n \ = \ n\,{\big /}\:\binom{n}{3} \ = \ \frac{6}{(n-1)(n-2)} .
\end{equation}

Figure 2a shows this function for \textit{n} = 500 together with the
result of a simulation made with one million random permutations of
\textit{n} pairwise different elements.

\begin{figure}[t]
\input{figure2}

{\small Figure 2. Simulation results and model functions for $p_n$
(\textit{n} = 500). Parameter \textit{m} is the sample size. Frequencies
of indices of partitioning elements are counted in bins of width 10. The
ordinate scales correspond to frequencies, transformed according to
equation (9), within bins. Correspondingly, the model functions have been
multiplied by the bin width.}

\vspace{-0.1in}

\end{figure}

\pagebreak
With this method, a selection of the partitioning element is not free
of charge. For $d_n$, \textit{n} $\ge$ 3, we have
\begin{equation}
 d_n(2) \ = \ \frac{1}{3}, \qquad \ \ \ d_n(3) \ = \ \frac{2}{3}, \qquad
 \ \ \ d_n(j) \ = \ 0 \ \ \mbox{otherwise}.
\end{equation}

To save the convolution with this function, we used the approximation
\begin{equation}
   d_n(3) \ = \ 1, \qquad \ \ \ d_n(j) \ = \ 0 \ \ \mbox{otherwise}
\end{equation}
instead, i.\,e. we assumed always three comparisons for the selection
process.

\subsection*{\textnormal{\textsl{4.3 \ Model 3: Median of three medians}}}

In their article on a new implementation of the C library function
\verb+qsort+, Bentley and McIlroy \cite{bent93} showed the usage of
selection method ``median of three medians, each of three elements''
(or simply ``median of three medians'') for array sizes \textit{n}
above some threshold value. A set of nine sample elements is divided
into three equally-sized subsets. The median within each subset is
determined, and the median of these three medians is used for
partitioning.

The probability function for the final position of partitioning elements,
resulting from this method, has been given by Durand \cite{durand}.
\pagebreak
Rewritten according to our conventions, it reads
($n \ge 9, \ 0 \le i < n$):
\begin{eqnarray}
p_n(i) \!\! & = \ b_n \ {\Big (} \!\!\!\! & 3 \prod_{j=0}^2 (i-j) \
\prod_{j=0}^4 (n-1-i-j) + 10 \prod_{j=0}^3 (i-j) (n-1-i-j) \nonumber \\
& & + \ 3 \prod_{j=0}^4 (i-j) \ \prod_{j=0}^2 (n-1-i-j) \ {\Big )} ,
\end{eqnarray}
where
\begin{equation}
b_n \ = \ 36\,{\big /} \prod_{j=0}^7 (n-1-j).
\end{equation}

In our computations, we used an approximation of this function instead,
mainly for consistency with those in models 4 and 5, namely
\begin{equation}
p_n(i) \ = \ a_n \ i (i-1) (i-2) (n-1-i) (n-2-i) (n-3-i), \\
\end{equation}
where factor $a_n$ was computed numerically. This is the simplest
polynomial that has the correct zeros at index positions 0, 1, 2,
$n$$-$1, $n$$-$2, $n$$-$3. The polynomial given in equation (23) has two
additional zeros at argument values
\begin{equation}
  \frac{n-1}{2} \ \pm \ \frac{1}{2} \sqrt{4 n^2 - 59 n + 217} .
\end{equation}
Our approximation slightly underestimates the concentration effect to
the middle of the array as can be seen in Figure 2b, where it is shown
together with the result of a simulation, again made with one million
random permutations.

The determination of the four medians of three may be interpreted
as a sequence of Bernoulli trials, each of them with probability $1/3$
for the ``success'' of needing only two comparisons. The corresponding
binomial distribution, transformed to the distribution of the total
number of comparisons for the selection, leads to
\begin{equation}
 d_n(j) \ = \ \binom{4}{j-8}\,\left({\frac{1}{3}}\right)^{12-j}
 \left({\frac{2}{3}}\right)^{j-8} \ \ \ \mbox{for} \ \ 8 \le j \le 12,
 \ \ \ 0 \ \ \mbox{otherwise}.
\end{equation}

Again, we avoided the convolution with this function for efficiency
reasons and assumed the maximum value of 12 comparisons for each
selection, in accordance with the principle given at the beginning of
this section.

\subsection*{\textnormal{\textsl{4.4 \ Model 4: Recursive median of
three medians}}}

The selection method of this model (``recursive median of medians'' for
short) is an extension of the methods from models 2 and 3 which suggests
itself. It contains these two as special cases.
For a given sample of \textit{m} elements, \textit{m} a power of 3 and
$\ge 3$, the method may be formulated as a recursive algorithm:

 \begin{lquote}

 {\sl Recursion basis:}

 If \textit{m} = 3, determine the median of the three elements.

 \medskip

{\sl Recursive part (if \textit{m} $\ge$ \textup{9}):}

 Divide the set of sample elements into three subsets, each of size
 \textit{m}$/$3. Determine the recursive median of medians element for
 each of these subsets by recursion. Then determine the median of these
 three elements.

\end{lquote}

If the sample is drawn from a set of pairwise different elements, there
must always be
\begin{equation}
  i_{\mbox{\scriptsize min}} \ = \ 2^k-1, \qquad\quad \mbox{where} \ \
  k \ = \ \log_3 m,
\end{equation}
elements less than and $i_{\mbox{\scriptsize min}}$ elements greater than
the partitioning element.
So we can infer
\begin{equation}
   p_n(i) \ = \ 0 \qquad \mbox{for} \ \ 0 \le i <
 i_{\mbox{\scriptsize min}} \ \ \mbox{and for} \ \
 n-i_{\mbox{\scriptsize min}} \le i < n.
\end{equation}

As an approximation for $p_n$ we used the simplest polynomial that has
zeros at the correct index positions $i$, this is
\begin{equation}
  p_n(i) \ = \ a_n \prod_{j=0}^{i_{\mbox{\scriptsize min}}-1} (i-j)
  (n-1-i-j).
\end{equation}
For \textit{m} $\ge$ 9, factor $a_n$ was computed numerically.

Figures 2c and 2d display this approximation for \textit{m} = 27 and
\textit{m} = 81, respectively, again together with results of simulations
with one million random permutations each. The sequence of Figures 2a
to 2d illustrates the increasing concentration effect towards the middle
of the array if \textit{m} is increased. It also shows that our
approximations increasingly underestimate this effect with increasing
\textit{m}.

The number of median of three determinations needed for a selection is
\begin{equation}
   t \ = \ \frac{m-1}{2}.
\end{equation}
So the examination that led to equation (27) can be generalized to
sequences of $t$ Bernoulli trials, which results in
\begin{equation}
 d_n(j) \ = \ \binom{t}{j-2t}\,\left({\frac{1}{3}}\right)^{3t-j}
 \left({\frac{2}{3}}\right)^{j-2t} \ \ \ \mbox{for} \ \ 2t \le j \le 3t,
 \ \ \ 0 \ \ \mbox{otherwise}.
\end{equation}
Again, we used the maximum value of $3t$ comparisons instead of
computing the convolution with this function.

\smallskip

To support the formation of well balanced recursion trees, we wish
that sample sizes increase \linebreak with \textit{n}. So we choose
\textit{m} as the greatest power of 3 satisfying
\begin{equation}
  q_{\mbox{\scriptsize min}} \ \le \ \frac{n}{m},
\end{equation}
where $q_{\mbox{\scriptsize min}}$ is an integer parameter $\ge 1$.
This leads to a number of comparisons needed for the selection process
that is bounded as
\begin{equation}
 \widehat{C}^{\mbox{\scriptsize {\,select}}}_n \ \le \
   \frac{3}{2} \left(\frac{n}{q_{\mbox{\scriptsize min}}} - 1\right)
   \qquad\mbox{for} \ \ n \ge 3 q_{\mbox{\scriptsize min}},
\end{equation}
i.\,e. by a linear function in \textit{n}. Thus with this method
Quicksort maintains its $O(n \log n)$ complexity for the average case.

An increase of the sample size also helps to prevent that ``non-random''
data, i.\,e. sequences already possessing some order, are preferred
candidates for bad cases.

\smallskip

In the following, the term ``recursive median of medians'' will be used
for this adaptive version where the sample size depends on \textit{n},
unless specifically stated otherwise.

\smallskip

It should be mentioned that there exists a method to determine the
$k$th smallest element---and thus also the median---of \textit{n}
elements that has worst case time complexity $O(n)$ (Blum et al.
\cite{blum73}). Its high cost, however, does not recommend its use
within Quicksort.

\pagebreak
\subsection*{\textnormal{\textsl{4.5 \ Model 5: Recursive median of
medians plus insertion sort}}}

Already in 1962, Hillmore \cite{hillm} remarked that the performance of
Quicksort could be improved if \textit{n} = 2 was treated as a basis
case of the recursive algorithm. Hibbard \cite{hibb63} suggested to
extend the basis further and use Shellsort for small arrays. In the
program version of Singleton \cite{singl}, Shellsort was replaced by
insertion sort, which has become standard since then. For this method
(``sorting by straight insertion''), the frequency distributions of
comparison may be computed from the recurrence
\begin{eqnarray}
 f_0(0) \!\! & = \!\! & f_1(0) \ = \ 1, \\
 f_0(j) \!\! & = \!\! & f_1(j) \ = \ 0  \qquad\mbox{for} \ \, j > 0, \\
 f_n \!\! & = \!\! & \sum_{k=1}^{n-1}\,(1+\delta_{n-1}(k))\;\delta_k
      \ast f_{n-1} \ \qquad\mbox{for} \ \, n > 1.      \label{rec4}
\end{eqnarray}

The average number of comparisons needed by insertion sort is
\begin{equation}
 \overline{C}_n \ = \ \frac{n(n+3)}{4} - H_n,
\end{equation}
 where $H_n$ is the \textit{n}th harmonic number $\sum_{k=1}^n k^{-1}$.

For the maximum number of comparisons we have
\begin{equation}
 \widehat{C}_n \ = \ \frac{n(n-1)}{2}.
\end{equation}

In model 5, we extended model 4 so that equation (\ref{rec4}) holds for
values of \textit{n} up to a parameter value
$n^{\mbox{\scriptsize b}}_{\mbox{\scriptsize max}}$ and equation
(\ref{rec1}) for $n > n^{\mbox{\scriptsize b}}_{\mbox{\scriptsize max}}$.
It turned out that up to \textit{n} = 9, the maximum number of
comparisons needed for insertion sort is not greater than the respective
value for models 2 to 4. So for our computations we used
$n^{\mbox{\scriptsize b}}_{\mbox{\scriptsize max}}$ = 9 for model 5.

\section*{\textnormal{5. \ Technical details of computations}}

Most of our numerical computations were done on a Sun Fire X4270 M2
server machine with x86 architecture, running under Solaris 10.
The recurrence equations for the average numbers (\ref{rec2}) and
maximum numbers (\ref{rec3}) of comparisons  were transformed into
recursive programs---written in C---with dynamic programming. Thus most
invocations of the respective recursive functions to get a result simply
lead to a table lookup.

Our main development task was an implementation to solve recurrence
(\ref{rec1}) (plus recurrence (\ref{rec4}) in the case of model 5) for
the frequency distributions of comparisons. During program execution, for
all distribution functions $f_k$ computed so far the frequency values
within the functions' supports are stored in a large array accompanied
by an integer array containing limits of the supports and pointers.
Thus every distribution had to be computed only once. Wherever possible,
use has been made of symmetries in recurrence equations (\ref{rec1}),
(\ref{rec2}), and (\ref{rec3}).

As element type for the tables needed to implement dynamic programming
we used the extended precision floating point type (\verb+long+
\verb+double+ in C). On the x86 machine, it has a relative machine
accuracy of \xpten{-19} and a maximum value of about \xpten{4932}. We
took care to do the summations in a way that rounding errors are kept
small. Crosscheck runs on a Sun Fire 280R machine with UltraSPARC~III
processors and a relative machine accuracy of \xnum{2}{-34} for type
\verb+long+ \verb+double+ did not show any difference in the results as
printed with at least four significant decimal digits.

Unfortunately, application of the convolution theorem with fast Fourier
transform works only for rather small values of \textit{n}. For example,
with model 1 and
\textit{n} $\gtrsim$ 40 on the x86 machine and \textit{n} $\gtrsim$ 53
on the SPARC machine, rounding errors in Fourier coefficients lead to
results for the probability of bad cases which differ from those computed
using equation (12) directly. With increasing \textit{n}, the tails of
distributions get more and more transformed into pure rounding noise
if the convolution theorem is applied.

\section*{\textnormal{6. \ Results}}

We first present numerical results for worst and average cases,
i.\,e. results that could be obtained without the need to compute
frequency distributions of comparisons. Afterwards, we show results for
standard deviations of comparisons and probabilities of bad cases.

\begin{figure}[tb]
{\parindent0em \setlength{\unitlength}{1.0cm} \begin{picture}(14.34,7.65)
\put(2.92,1){\framebox(8.5,6){}}
\qbezier(2.92,1)(3.77,1.001)(4.62,1.219)
\qbezier(4.62,1.219)(5.47,1.438)(6.32,1.875)
\qbezier(6.32,1.875)(7.17,2.312)(8.02,2.967)
\qbezier(8.02,2.967)(8.87,3.622)(9.72,4.495)
\qbezier(9.72,4.495)(10.57,5.369)(11.42,6.46)
\small \put(2.92,1){\line(0,-1){0.164}}
\put(2.92,0.676){\makebox(0,0)[t]{0}}
\put(3.77,1){\line(0,-1){0.107}}
\put(4.62,1){\line(0,-1){0.164}}
\put(4.62,0.676){\makebox(0,0)[t]{200}}
\put(5.47,1){\line(0,-1){0.107}}
\put(6.32,1){\line(0,-1){0.164}}
\put(6.32,0.676){\makebox(0,0)[t]{400}}
\put(7.17,1){\line(0,-1){0.107}}
\put(8.02,1){\line(0,-1){0.164}}
\put(8.02,0.676){\makebox(0,0)[t]{600}}
\put(8.87,1){\line(0,-1){0.107}}
\put(9.72,1){\line(0,-1){0.164}}
\put(9.72,0.676){\makebox(0,0)[t]{800}}
\put(10.57,1){\line(0,-1){0.107}}
\put(11.42,1){\line(0,-1){0.164}}
\put(11.42,0.676){\makebox(0,0)[t]{1000}}
\put(3.77,7){\line(0,-1){0.107}}
\put(4.62,7){\line(0,-1){0.164}}
\put(5.47,7){\line(0,-1){0.107}}
\put(6.32,7){\line(0,-1){0.164}}
\put(7.17,7){\line(0,-1){0.107}}
\put(8.02,7){\line(0,-1){0.164}}
\put(8.87,7){\line(0,-1){0.107}}
\put(9.72,7){\line(0,-1){0.164}}
\put(10.57,7){\line(0,-1){0.107}}
\put(2.76,1){\makebox(0,0)[r]{0}}
\put(2.92,1.545){\line(1,0){0.107}}
\put(2.92,2.091){\line(1,0){0.164}}
\put(2.76,2.091){\makebox(0,0)[r]{100}}
\put(2.92,2.636){\line(1,0){0.107}}
\put(2.92,3.182){\line(1,0){0.164}}
\put(2.76,3.182){\makebox(0,0)[r]{200}}
\put(2.92,3.727){\line(1,0){0.107}}
\put(2.92,4.273){\line(1,0){0.164}}
\put(2.76,4.273){\makebox(0,0)[r]{300}}
\put(2.92,4.818){\line(1,0){0.107}}
\put(2.92,5.364){\line(1,0){0.164}}
\put(2.76,5.364){\makebox(0,0)[r]{400}}
\put(2.92,5.909){\line(1,0){0.107}}
\put(2.92,6.455){\line(1,0){0.164}}
\put(2.76,6.455){\makebox(0,0)[r]{500}}
\put(11.42,1.545){\line(-1,0){0.107}}
\put(11.42,2.091){\line(-1,0){0.164}}
\put(11.42,2.636){\line(-1,0){0.107}}
\put(11.42,3.182){\line(-1,0){0.164}}
\put(11.42,3.727){\line(-1,0){0.107}}
\put(11.42,4.273){\line(-1,0){0.164}}
\put(11.42,4.818){\line(-1,0){0.107}}
\put(11.42,5.364){\line(-1,0){0.164}}
\put(11.42,5.909){\line(-1,0){0.107}}
\put(11.42,6.455){\line(-1,0){0.164}}
\normalsize \qbezier(2.92,1)(3.77,1.002)(4.62,1.113)
\qbezier(4.62,1.113)(5.47,1.225)(6.32,1.445)
\qbezier(6.32,1.445)(7.17,1.665)(8.02,1.995)
\qbezier(8.02,1.995)(8.87,2.324)(9.72,2.763)
\qbezier(9.72,2.763)(10.57,3.201)(11.42,3.749)
\qbezier(2.92,1)(3.77,1.006)(4.62,1.067)
\qbezier(4.62,1.067)(5.472,1.127)(6.32,1.241)
\qbezier(6.32,1.241)(7.169,1.356)(8.02,1.525)
\qbezier(8.02,1.525)(8.87,1.694)(9.72,1.918)
\qbezier(9.72,1.918)(10.57,2.141)(11.42,2.419)
\qbezier(2.92,1)(3.77,1.024)(4.62,1.073)
\qbezier(4.62,1.073)(5.324,1.114)(6.32,1.192)
\qbezier(6.32,1.192)(7.372,1.274)(8.02,1.346)
\qbezier(8.02,1.346)(8.87,1.451)(9.72,1.556)
\qbezier(9.72,1.556)(10.57,1.654)(11.42,1.711)
\qbezier(2.92,1)(3.77,1.011)(4.62,1.053)
\qbezier(4.62,1.053)(4.631,1.054)(6.32,1.154)
\qbezier(6.32,1.154)(7.17,1.202)(8.02,1.251)
\qbezier(8.02,1.251)(8.791,1.299)(9.72,1.375)
\qbezier(9.72,1.375)(10.57,1.445)(11.42,1.527)
\qbezier(2.92,1)(3.77,1.011)(4.62,1.052)
\qbezier(4.62,1.052)(4.673,1.055)(6.32,1.152)
\qbezier(6.32,1.152)(7.17,1.2)(8.02,1.249)
\qbezier(8.02,1.249)(8.791,1.297)(9.72,1.373)
\qbezier(9.72,1.373)(10.57,1.443)(11.42,1.525)
\put(2.92,7.36){\makebox(0,0)[bl]{Maximum numbers of comparisons}}
\small \put(7.17,0.31){\makebox(0,0)[t]{\textit{n}}}
\put(1.9,4){\makebox(0,0)[r]{$\widehat{C}_n/$\xpten{3}}}
\put(10.145,5.14){\makebox(0,0)[b]{1}}
\put(10.145,3.1){\makebox(0,0)[b]{2}}
\put(10.145,2.122){\makebox(0,0)[b]{3}}
\put(10.145,1.678){\makebox(0,0)[b]{4a}}
\put(10.145,1.12){\makebox(0,0)[b]{4b \ 5}}
\put(3.94,6.4){\makebox(0,0)[tr]{1:}}
\put(4.11,6.4){\makebox(0,0)[tl]{Model 1}}
\put(3.94,6.01){\makebox(0,0)[tr]{2:}}
\put(4.11,6.01){\makebox(0,0)[tl]{Model 2}}
\put(3.94,5.62){\makebox(0,0)[tr]{3:}}
\put(4.11,5.62){\makebox(0,0)[tl]{Model 3, $q_{\mbox{\scriptsize min}}$ = 5}}
\put(3.94,5.23){\makebox(0,0)[tr]{4a:}}
\put(4.11,5.23){\makebox(0,0)[tl]{Model 4, $q_{\mbox{\scriptsize min}}$ = 10}}
\put(3.94,4.84){\makebox(0,0)[tr]{4b:}}
\put(4.11,4.84){\makebox(0,0)[tl]{Model 4, $q_{\mbox{\scriptsize min}}$ = 5}}
\put(3.94,4.45){\makebox(0,0)[tr]{5:}}
\put(4.11,4.54){\makebox(0,0)[tl]{Model 5, $q_{\mbox{\scriptsize min}}$ = 5, $n^{\mbox{\scriptsize b}}_{\mbox{\scriptsize max}}$ = 9}}
\end{picture}
}
{\small Figure 3. Maximum numbers of comparisons.
Curves 4b and 5 are visually indistinguishable.}

\vspace{0.2in}

{\parindent0em \setlength{\unitlength}{1.0cm} \begin{picture}(14.34,7.65)
\put(2.92,1){\framebox(8.5,6){}}
\qbezier(2.92,1)(3.132,1.032)(3.345,1.093)
\qbezier(3.345,1.093)(3.529,1.147)(3.77,1.241)
\qbezier(3.77,1.241)(3.981,1.324)(4.195,1.421)
\qbezier(4.195,1.421)(4.399,1.513)(4.62,1.624)
\qbezier(4.62,1.624)(4.828,1.729)(5.045,1.847)
\qbezier(5.045,1.847)(5.253,1.961)(5.47,2.087)
\qbezier(5.47,2.087)(5.679,2.21)(5.895,2.343)
\qbezier(5.895,2.343)(6.105,2.473)(6.32,2.613)
\qbezier(6.32,2.613)(6.53,2.749)(6.745,2.895)
\qbezier(6.745,2.895)(6.955,3.038)(7.17,3.189)
\qbezier(7.17,3.189)(7.38,3.338)(7.595,3.495)
\qbezier(7.595,3.495)(7.806,3.649)(8.02,3.811)
\qbezier(8.02,3.811)(8.231,3.97)(8.445,4.136)
\qbezier(8.445,4.136)(8.656,4.3)(8.87,4.472)
\qbezier(8.87,4.472)(9.081,4.64)(9.295,4.816)
\qbezier(9.295,4.816)(9.506,4.989)(9.72,5.169)
\qbezier(9.72,5.169)(9.931,5.346)(10.145,5.53)
\qbezier(10.145,5.53)(10.356,5.711)(10.57,5.898)
\qbezier(10.57,5.898)(10.781,6.084)(10.995,6.275)
\qbezier(10.995,6.275)(11.207,6.465)(11.42,6.659)
\small \put(2.92,0.84){\makebox(0,0)[t]{0}}
\put(3.6,1){\line(0,1){0.107}}
\put(4.28,1){\line(0,1){0.107}}
\put(4.96,1){\line(0,1){0.107}}
\put(5.64,1){\line(0,1){0.107}}
\put(6.32,1){\line(0,1){0.164}}
\put(6.32,0.84){\makebox(0,0)[t]{10}}
\put(7,1){\line(0,1){0.107}}
\put(7.68,1){\line(0,1){0.107}}
\put(8.36,1){\line(0,1){0.107}}
\put(9.04,1){\line(0,1){0.107}}
\put(9.72,1){\line(0,1){0.164}}
\put(9.72,0.84){\makebox(0,0)[t]{20}}
\put(10.4,1){\line(0,1){0.107}}
\put(11.08,1){\line(0,1){0.107}}
\put(3.6,7){\line(0,-1){0.107}}
\put(4.28,7){\line(0,-1){0.107}}
\put(4.96,7){\line(0,-1){0.107}}
\put(5.64,7){\line(0,-1){0.107}}
\put(6.32,7){\line(0,-1){0.164}}
\put(7,7){\line(0,-1){0.107}}
\put(7.68,7){\line(0,-1){0.107}}
\put(8.36,7){\line(0,-1){0.107}}
\put(9.04,7){\line(0,-1){0.107}}
\put(9.72,7){\line(0,-1){0.164}}
\put(10.4,7){\line(0,-1){0.107}}
\put(11.08,7){\line(0,-1){0.107}}
\put(2.76,1){\makebox(0,0)[r]{0}}
\put(2.92,1.545){\line(1,0){0.107}}
\put(2.92,2.091){\line(1,0){0.164}}
\put(2.76,2.091){\makebox(0,0)[r]{10}}
\put(2.92,2.636){\line(1,0){0.107}}
\put(2.92,3.182){\line(1,0){0.164}}
\put(2.76,3.182){\makebox(0,0)[r]{20}}
\put(2.92,3.727){\line(1,0){0.107}}
\put(2.92,4.273){\line(1,0){0.164}}
\put(2.76,4.273){\makebox(0,0)[r]{30}}
\put(2.92,4.818){\line(1,0){0.107}}
\put(2.92,5.364){\line(1,0){0.164}}
\put(2.76,5.364){\makebox(0,0)[r]{40}}
\put(2.92,5.909){\line(1,0){0.107}}
\put(2.92,6.455){\line(1,0){0.164}}
\put(2.76,6.455){\makebox(0,0)[r]{50}}
\put(11.42,1.545){\line(-1,0){0.107}}
\put(11.42,2.091){\line(-1,0){0.164}}
\put(11.42,2.636){\line(-1,0){0.107}}
\put(11.42,3.182){\line(-1,0){0.164}}
\put(11.42,3.727){\line(-1,0){0.107}}
\put(11.42,4.273){\line(-1,0){0.164}}
\put(11.42,4.818){\line(-1,0){0.107}}
\put(11.42,5.364){\line(-1,0){0.164}}
\put(11.42,5.909){\line(-1,0){0.107}}
\put(11.42,6.455){\line(-1,0){0.164}}
\scriptsize \put(2.92,1){\makebox(0,0){$\bullet$}}
\put(3.09,1.025){\makebox(0,0){$\bullet$}}
\put(3.26,1.067){\makebox(0,0){$\bullet$}}
\put(3.43,1.111){\makebox(0,0){$\bullet$}}
\put(3.6,1.168){\makebox(0,0){$\bullet$}}
\put(3.77,1.238){\makebox(0,0){$\bullet$}}
\put(3.94,1.299){\makebox(0,0){$\bullet$}}
\put(4.11,1.358){\makebox(0,0){$\bullet$}}
\put(4.28,1.423){\makebox(0,0){$\bullet$}}
\put(4.45,1.495){\makebox(0,0){$\bullet$}}
\put(4.62,1.573){\makebox(0,0){$\bullet$}}
\put(4.79,1.659){\makebox(0,0){$\bullet$}}
\put(4.96,1.75){\makebox(0,0){$\bullet$}}
\put(5.13,1.849){\makebox(0,0){$\bullet$}}
\put(5.3,1.954){\makebox(0,0){$\bullet$}}
\put(5.47,2.066){\makebox(0,0){$\bullet$}}
\put(5.64,2.181){\makebox(0,0){$\bullet$}}
\put(5.81,2.258){\makebox(0,0){$\bullet$}}
\put(5.98,2.338){\makebox(0,0){$\bullet$}}
\put(6.15,2.421){\makebox(0,0){$\bullet$}}
\put(6.32,2.508){\makebox(0,0){$\bullet$}}
\put(6.49,2.598){\makebox(0,0){$\bullet$}}
\put(6.66,2.691){\makebox(0,0){$\bullet$}}
\put(6.83,2.788){\makebox(0,0){$\bullet$}}
\put(7,2.888){\makebox(0,0){$\bullet$}}
\put(7.17,2.991){\makebox(0,0){$\bullet$}}
\put(7.34,3.098){\makebox(0,0){$\bullet$}}
\put(7.51,3.207){\makebox(0,0){$\bullet$}}
\put(7.68,3.321){\makebox(0,0){$\bullet$}}
\put(7.85,3.437){\makebox(0,0){$\bullet$}}
\put(8.02,3.557){\makebox(0,0){$\bullet$}}
\put(8.19,3.681){\makebox(0,0){$\bullet$}}
\put(8.36,3.807){\makebox(0,0){$\bullet$}}
\put(8.53,3.937){\makebox(0,0){$\bullet$}}
\put(8.7,4.07){\makebox(0,0){$\bullet$}}
\put(8.87,4.207){\makebox(0,0){$\bullet$}}
\put(9.04,4.347){\makebox(0,0){$\bullet$}}
\put(9.21,4.49){\makebox(0,0){$\bullet$}}
\put(9.38,4.636){\makebox(0,0){$\bullet$}}
\put(9.55,4.786){\makebox(0,0){$\bullet$}}
\put(9.72,4.939){\makebox(0,0){$\bullet$}}
\put(9.89,5.096){\makebox(0,0){$\bullet$}}
\put(10.06,5.256){\makebox(0,0){$\bullet$}}
\put(10.23,5.419){\makebox(0,0){$\bullet$}}
\put(10.4,5.585){\makebox(0,0){$\bullet$}}
\put(10.57,5.755){\makebox(0,0){$\bullet$}}
\put(10.74,5.928){\makebox(0,0){$\bullet$}}
\put(10.91,6.105){\makebox(0,0){$\bullet$}}
\put(11.08,6.272){\makebox(0,0){$\bullet$}}
\put(11.25,6.384){\makebox(0,0){$\bullet$}}
\put(11.42,6.498){\makebox(0,0){$\bullet$}}
\small \linethickness{0.22mm}
\qbezier[13](5.63,1)(5.63,1.591)(5.63,2.182)
\thinlines
\put(5.715,1.591){\makebox(0,0)[l]{\textit{n} = $\mbox{3}^{13}q_{\mbox{\scriptsize min}}$}}
\linethickness{0.22mm}
\qbezier[54](11.051,1)(11.051,3.663)(11.051,6.325)
\thinlines
\put(10.966,3.663){\makebox(0,0)[r]{\textit{n} = $\mbox{3}^{14}q_{\mbox{\scriptsize min}}$}}
\normalsize \put(2.92,7.36){\makebox(0,0)[bl]{Maximum numbers of comparisons}}
\small \put(7.17,0.46){\makebox(0,0)[t]{\textit{n}\,/\xpten{6}}}
\put(2.07,4){\makebox(0,0)[r]{$\widehat{C}_n/$\xpten{9}}}
\scriptsize \put(3.982,6.1){\makebox(0,0){$\bullet$}} \small
\put(4.705,6.1){\makebox(0,0)[l]%
{Model 5, $q_{\mbox{\scriptsize min}}$ = 5, $n^{\mbox{\scriptsize b}}_{\mbox{\scriptsize max}}$ = 9}}
\put(3.558,5.56){\line(1,0){0.85}}
\put(4.705,5.56){\makebox(0,0)[l]{3.8\,$n^{1.37}$}}
\end{picture}
}
{\small Figure 4. Maximum numbers of comparisons for model 5, displayed
for values of \textit{n} up to 25 millions. Two thresholds for sample
sizes are indicated by dotted lines. Additionally, the figure shows
the graph of a function that may be an asymptotical upper bound for
$\widehat{C}_n$.}
\end{figure}

Parameter name $q_{\mbox{\scriptsize min}}$, defined for models 4 and 5
(equation (33)), will also be used for model 3, meaning here that
selection method median of three medians is used for arrays of size
$n \ge 9\,q_{\mbox{\scriptsize min}}$, and method median of three
elements below this threshold.

\subsection*{\textnormal{\textsl{6.1 \ Worst case}}}

Figure 3 shows the maximum numbers of comparisons for several models and
parameters up to \textit{n}~=~1000, computed from the recurrence
equation (\ref{rec3}).

If the recursive median of medians method with a fixed sample size is
used for selection of the partitioning element, the worst case
complexity of Quicksort is $O(n^2)$. In particular:
for a sample size $m = 3^k$, $k \ge 0$, it is theoretically expected that
the maximum number of comparisons is
\begin{equation}
 \widehat{C}_n \ = \ 2^{-(k+1)}\,n^2 \ + \ \mbox{lower order}.
\end{equation}

Curves 1, 2, and 3 in Figure 3 can be regarded as compatible with the
expectation of leading terms $n^2/2$, $n^2/4$, $n^2/8$, respectively.
For models 4 and 5, however, the increase of the curves appears much
slower than $n^2$. This is not surprising, because the sample size is
stepwise increasing with \textit{n}. The fact that curves 4b and 5
(models 4 and 5 with the same value for $q_{\mbox{\scriptsize min}}$)
are visually indistinguishable shows that the influence of using
insertion sort for small arrays is negligible for the worst case, as
could be expected from our choice of
$n^{\mbox{\scriptsize b}}_{\mbox{\scriptsize max}}$ (Section 4.5).

\smallskip

We also used McIlroy's ``killer adversary'' \cite{mcilr} to construct
bad cases for our Quicksort versions. The adversary program was applied
to models 1 to 5, with $q_{\mbox{\scriptsize min}}=5$ for models 3 to 5
and $n^{\mbox{\scriptsize b}}_{\mbox{\scriptsize max}}=9$ for model 5.
The following numbers of comparisons were determined:
{
\setlength{\tabcolsep}{0.12in}
\begin{center}
\begin{tabular}{rr@{\,$\cdot$\,}lr@{\,$\cdot$\,}l
r@{\,$\cdot$\,}lr@{\,$\cdot$\,}lr@{\,$\cdot$\,}l}
\multicolumn{1}{c}{\textit{n}} & \multicolumn{2}{c}{Model 1} &
\multicolumn{2}{c}{Model 2} & \multicolumn{2}{c}{Model 3} &
\multicolumn{2}{c}{Model 4} & \multicolumn{2}{c}{Model 5} \\ \hline
\multicolumn{2}{c}{ } \\[-2.0ex]
100,000 & 2.500 & \xpten{9} & 2.500 & \xpten{9}
 & 8.338 & \xpten{8} & 2.848 & \xpten{6} & 2.768 & \xpten{6} \\
200,000 & 1.000 & \xpten{10} & 1.000 & \xpten{10}
 & 3.334 & \xpten{9} & 5.922 & \xpten{6} & 5.760 & \xpten{6} \\
500,000 & 6.250 & \xpten{10} & 6.250 & \xpten{10}
 & 2.084 & \xpten{10} & 1.515 & \xpten{7} & 1.474 & \xpten{7} \\
1,000,000 & 2.500 & \xpten{11} & 2.500 & \xpten{11}
 & 8.334 & \xpten{10} & 3.100 & \xpten{7} & 3.016 & \xpten{7}
\end{tabular}
\end{center}
}
These data, too, suggest that for models 4 and 5 the increase is much
slower than the $O(n^2)$ behavior of models 1 to 3, even though the
adversary program assumes $O(1)$ complexity for the selection of the
partitioning element whereas it is actually $O(n)$ for those two models
(equation (34)).

\smallskip

To get an estimate for the complexity of our recursive median of medians
models (4 and 5), we computed the maximum numbers of comparisons for
model 5 up to \textit{n} = 25 millions. The result is shown in Figure 4.
In addition to our data, we display a function
\begin{equation}
 b(n) \ = \ 3.8 \ n^{1.37}
\end{equation}
which was determined such that it might be an asymptotical upper bound
for $\widehat{C}_n$.

\subsection*{\textnormal{\textsl{6.2 \ Average case}}}

Average numbers of comparisons were computed from the recurrence
equation (\ref{rec2}). To keep Figure~5 readable, results are shown only
for four of the six model/parameter combinations that were displayed in
Figure 3. Additionally, we list average numbers for all models and for
several values of \textit{n} up to 10,000 (model identifiers
corresponding to the numbering of curves in Figure 3):

\begin{figure}[tb]
{\parindent0em \setlength{\unitlength}{1.0cm} \begin{picture}(14.34,7.65)
\put(2.92,1){\framebox(8.5,6){}}
\qbezier(2.92,1)(3.77,1.329)(4.62,1.815)
\qbezier(4.62,1.815)(5.276,2.19)(6.32,2.901)
\qbezier(6.32,2.901)(7.205,3.504)(8.02,4.092)
\qbezier(8.02,4.092)(8.799,4.654)(9.72,5.35)
\qbezier(9.72,5.35)(10.57,5.994)(11.42,6.66)
\small \put(2.92,0.84){\makebox(0,0)[t]{0}}
\put(3.77,1){\line(0,1){0.107}}
\put(4.62,1){\line(0,1){0.164}}
\put(4.62,0.84){\makebox(0,0)[t]{200}}
\put(5.47,1){\line(0,1){0.107}}
\put(6.32,1){\line(0,1){0.164}}
\put(6.32,0.84){\makebox(0,0)[t]{400}}
\put(7.17,1){\line(0,1){0.107}}
\put(8.02,1){\line(0,1){0.164}}
\put(8.02,0.84){\makebox(0,0)[t]{600}}
\put(8.87,1){\line(0,1){0.107}}
\put(9.72,1){\line(0,1){0.164}}
\put(9.72,0.84){\makebox(0,0)[t]{800}}
\put(10.57,1){\line(0,1){0.107}}
\put(11.42,0.84){\makebox(0,0)[t]{1000}}
\put(3.77,7){\line(0,-1){0.107}}
\put(4.62,7){\line(0,-1){0.164}}
\put(5.47,7){\line(0,-1){0.107}}
\put(6.32,7){\line(0,-1){0.164}}
\put(7.17,7){\line(0,-1){0.107}}
\put(8.02,7){\line(0,-1){0.164}}
\put(8.87,7){\line(0,-1){0.107}}
\put(9.72,7){\line(0,-1){0.164}}
\put(10.57,7){\line(0,-1){0.107}}
\put(2.76,1){\makebox(0,0)[r]{0}}
\put(2.92,1.5){\line(1,0){0.107}}
\put(2.92,2){\line(1,0){0.164}}
\put(2.76,2){\makebox(0,0)[r]{2000}}
\put(2.92,2.5){\line(1,0){0.107}}
\put(2.92,3){\line(1,0){0.164}}
\put(2.76,3){\makebox(0,0)[r]{4000}}
\put(2.92,3.5){\line(1,0){0.107}}
\put(2.92,4){\line(1,0){0.164}}
\put(2.76,4){\makebox(0,0)[r]{6000}}
\put(2.92,4.5){\line(1,0){0.107}}
\put(2.92,5){\line(1,0){0.164}}
\put(2.76,5){\makebox(0,0)[r]{8000}}
\put(2.92,5.5){\line(1,0){0.107}}
\put(2.92,6){\line(1,0){0.164}}
\put(2.76,6){\makebox(0,0)[r]{10000}}
\put(2.92,6.5){\line(1,0){0.107}}
\put(2.76,7){\makebox(0,0)[r]{12000}}
\put(11.42,1.5){\line(-1,0){0.107}}
\put(11.42,2){\line(-1,0){0.164}}
\put(11.42,2.5){\line(-1,0){0.107}}
\put(11.42,3){\line(-1,0){0.164}}
\put(11.42,3.5){\line(-1,0){0.107}}
\put(11.42,4){\line(-1,0){0.164}}
\put(11.42,4.5){\line(-1,0){0.107}}
\put(11.42,5){\line(-1,0){0.164}}
\put(11.42,5.5){\line(-1,0){0.107}}
\put(11.42,6){\line(-1,0){0.164}}
\put(11.42,6.5){\line(-1,0){0.107}}
\normalsize \qbezier(2.92,1)(3.77,1.345)(4.62,1.816)
\qbezier(4.62,1.816)(5.272,2.178)(6.32,2.852)
\qbezier(6.32,2.852)(7.211,3.425)(8.02,3.97)
\qbezier(8.02,3.97)(8.798,4.494)(9.72,5.141)
\qbezier(9.72,5.141)(10.57,5.737)(11.42,6.352)
\qbezier(2.92,1)(3.77,1.351)(4.62,1.827)
\qbezier(4.62,1.827)(5.347,2.234)(6.32,2.879)
\qbezier(6.32,2.879)(7.034,3.352)(8.02,4.044)
\qbezier(8.02,4.044)(8.869,4.64)(9.72,5.254)
\qbezier(9.72,5.254)(10.57,5.868)(11.42,6.498)
\qbezier(2.92,1)(3.77,1.321)(4.62,1.766)
\qbezier(4.62,1.766)(5.347,2.147)(6.32,2.758)
\qbezier(6.32,2.758)(7.034,3.206)(8.02,3.863)
\qbezier(8.02,3.863)(8.869,4.429)(9.72,5.013)
\qbezier(9.72,5.013)(10.57,5.596)(11.42,6.197)
\put(2.92,7.36){\makebox(0,0)[bl]{Average numbers of comparisons}}
\small \put(7.17,0.4){\makebox(0,0)[t]{\textit{n}}}
\put(1.645,4){\makebox(0,0)[r]{$\overline{C}_n$}}
\put(10.357,5.836){\line(-1,1){0.36}}
\put(9.998,6.274){\makebox(0,0)[b]{1}}
\put(9.72,5.254){\line(-1,1){0.36}}
\put(9.36,5.692){\makebox(0,0)[b]{4b}}
\put(10.145,5.44){\line(1,-1){0.43}}
\put(10.575,4.932){\makebox(0,0)[t]{3}}
\put(9.465,4.838){\line(1,-1){0.36}}
\put(9.825,4.4){\makebox(0,0)[t]{5}}
\end{picture}
}
{\small Figure 5. Average numbers of comparisons.
Numbering of curves is the same as in Figure 3.}

\vspace{0.2in}

{\parindent0em \setlength{\unitlength}{1.0cm} \begin{picture}(14.34,7.65)
\put(2.92,1){\framebox(8.5,6){}}
\qbezier(2.92,1)(3.77,1.536)(4.62,2.12)
\qbezier(4.62,2.12)(5.08,2.436)(6.32,3.317)
\qbezier(6.32,3.317)(7.17,3.919)(8.02,4.522)
\qbezier(8.02,4.522)(8.61,4.94)(9.72,5.729)
\qbezier(9.72,5.729)(10.57,6.334)(11.42,6.939)
\small \put(2.92,0.84){\makebox(0,0)[t]{0}}
\put(3.77,1){\line(0,1){0.107}}
\put(4.62,1){\line(0,1){0.164}}
\put(4.62,0.84){\makebox(0,0)[t]{100}}
\put(5.47,1){\line(0,1){0.107}}
\put(6.32,1){\line(0,1){0.164}}
\put(6.32,0.84){\makebox(0,0)[t]{200}}
\put(7.17,1){\line(0,1){0.107}}
\put(8.02,1){\line(0,1){0.164}}
\put(8.02,0.84){\makebox(0,0)[t]{300}}
\put(8.87,1){\line(0,1){0.107}}
\put(9.72,1){\line(0,1){0.164}}
\put(9.72,0.84){\makebox(0,0)[t]{400}}
\put(10.57,1){\line(0,1){0.107}}
\put(11.42,0.84){\makebox(0,0)[t]{500}}
\put(3.77,7){\line(0,-1){0.107}}
\put(4.62,7){\line(0,-1){0.164}}
\put(5.47,7){\line(0,-1){0.107}}
\put(6.32,7){\line(0,-1){0.164}}
\put(7.17,7){\line(0,-1){0.107}}
\put(8.02,7){\line(0,-1){0.164}}
\put(8.87,7){\line(0,-1){0.107}}
\put(9.72,7){\line(0,-1){0.164}}
\put(10.57,7){\line(0,-1){0.107}}
\put(2.76,1){\makebox(0,0)[r]{0}}
\put(2.92,1.375){\line(1,0){0.107}}
\put(2.92,1.75){\line(1,0){0.107}}
\put(2.92,2.125){\line(1,0){0.107}}
\put(2.92,2.5){\line(1,0){0.107}}
\put(2.92,2.875){\line(1,0){0.164}}
\put(2.76,2.875){\makebox(0,0)[r]{100}}
\put(2.92,3.25){\line(1,0){0.107}}
\put(2.92,3.625){\line(1,0){0.107}}
\put(2.92,4){\line(1,0){0.107}}
\put(2.92,4.375){\line(1,0){0.107}}
\put(2.92,4.75){\line(1,0){0.164}}
\put(2.76,4.75){\makebox(0,0)[r]{200}}
\put(2.92,5.125){\line(1,0){0.107}}
\put(2.92,5.5){\line(1,0){0.107}}
\put(2.92,5.875){\line(1,0){0.107}}
\put(2.92,6.25){\line(1,0){0.107}}
\put(2.92,6.625){\line(1,0){0.164}}
\put(2.76,6.625){\makebox(0,0)[r]{300}}
\put(11.42,1.375){\line(-1,0){0.107}}
\put(11.42,1.75){\line(-1,0){0.107}}
\put(11.42,2.125){\line(-1,0){0.107}}
\put(11.42,2.5){\line(-1,0){0.107}}
\put(11.42,2.875){\line(-1,0){0.164}}
\put(11.42,3.25){\line(-1,0){0.107}}
\put(11.42,3.625){\line(-1,0){0.107}}
\put(11.42,4){\line(-1,0){0.107}}
\put(11.42,4.375){\line(-1,0){0.107}}
\put(11.42,4.75){\line(-1,0){0.164}}
\put(11.42,5.125){\line(-1,0){0.107}}
\put(11.42,5.5){\line(-1,0){0.107}}
\put(11.42,5.875){\line(-1,0){0.107}}
\put(11.42,6.25){\line(-1,0){0.107}}
\put(11.42,6.625){\line(-1,0){0.164}}
\qbezier(2.92,1)(3.77,1.265)(4.62,1.564)
\qbezier(4.62,1.564)(5.093,1.731)(6.32,2.185)
\qbezier(6.32,2.185)(7.17,2.498)(8.02,2.812)
\qbezier(8.02,2.812)(8.626,3.035)(9.72,3.441)
\qbezier(9.72,3.441)(10.57,3.756)(11.42,4.071)
\qbezier(2.92,1)(3.77,1.211)(4.62,1.381)
\qbezier(4.62,1.381)(4.925,1.441)(6.32,1.697)
\qbezier(6.32,1.697)(6.789,1.782)(8.02,2.016)
\qbezier(8.02,2.016)(8.87,2.176)(9.72,2.337)
\qbezier(9.72,2.337)(10.57,2.497)(11.42,2.659)
\qbezier(2.92,1)(3.77,1.252)(4.62,1.448)
\qbezier(4.62,1.448)(5.245,1.592)(6.32,1.799)
\qbezier(6.32,1.799)(6.907,1.913)(7.493,2.023)
\qbezier(7.493,2.023)(7.502,1.958)(7.51,1.894)
\qbezier(7.51,1.894)(7.765,1.932)(8.02,1.974)
\qbezier(8.02,1.974)(8.87,2.112)(9.72,2.251)
\qbezier(9.72,2.251)(10.57,2.373)(11.42,2.43)
\qbezier(2.92,1)(3.719,1.204)(4.62,1.381)
\qbezier(4.62,1.381)(4.909,1.437)(5.198,1.487)
\qbezier(5.198,1.487)(5.207,1.455)(5.215,1.424)
\qbezier(5.215,1.424)(5.768,1.566)(6.32,1.636)
\qbezier(6.32,1.636)(7.17,1.705)(8.02,1.774)
\qbezier(8.02,1.774)(8.321,1.802)(8.87,1.872)
\qbezier(8.87,1.872)(9.329,1.931)(9.788,1.992)
\qbezier(9.788,1.992)(9.796,1.96)(9.805,1.927)
\qbezier(9.805,1.927)(10.101,1.959)(10.57,1.998)
\qbezier(10.57,1.998)(10.995,2.033)(11.42,2.053)
\qbezier(2.92,1)(3.719,1.235)(4.62,1.42)
\qbezier(4.62,1.42)(4.909,1.479)(5.198,1.529)
\qbezier(5.198,1.529)(5.207,1.5)(5.215,1.471)
\qbezier(5.215,1.471)(5.768,1.611)(6.32,1.683)
\qbezier(6.32,1.683)(7.17,1.758)(8.02,1.832)
\qbezier(8.02,1.832)(8.321,1.861)(8.87,1.933)
\qbezier(8.87,1.933)(9.329,1.992)(9.788,2.054)
\qbezier(9.788,2.054)(9.796,2.023)(9.805,1.993)
\qbezier(9.805,1.993)(10.101,2.025)(10.57,2.066)
\qbezier(10.57,2.066)(10.995,2.102)(11.42,2.124)
\normalsize \put(2.92,7.36){\makebox(0,0)[bl]{Standard deviations of numbers of comparisons}}
\small \put(1.9,4){\makebox(0,0)[r]{$\sigma_{\mbox{\scriptsize \textit{n}}}$}}
\put(7.17,0.4){\makebox(0,0)[t]{\textit{n}}}
\put(10.145,6.148){\makebox(0,0)[b]{1}}
\put(10.145,3.682){\makebox(0,0)[b]{2}}
\put(10.145,2.5){\makebox(0,0)[b]{3}}
\put(10.995,2.17){\makebox(0,0)[b]{4a}}
\put(10.145,1.678){\makebox(0,0)[b]{4b}}
\put(10.74,2.08){\line(1,-1){0.36}}
\put(11.1,1.642){\makebox(0,0)[t]{5}}
\put(8.87,1.933){\line(1,-1){0.36}}
\put(9.23,1.495){\makebox(0,0)[t]{5}}
\end{picture}
}
{\small Figure 6. Standard deviations of numbers of comparison.
Numbering of curves is the same as in Figure~3.}

\vspace{-0.1in}

\end{figure}

{
\tabcolsep0.06in
\begin{center}
\begin{tabular}{crccrccrccrccrccrccrc}
\multicolumn{3}{c}{\textit{n}} &
\multicolumn{3}{c}{Model 1} &
\multicolumn{3}{c}{Model 2} &
\multicolumn{3}{c}{Model 3} &
\multicolumn{3}{c}{Model 4a} &
\multicolumn{3}{c}{Model 4b} &
\multicolumn{3}{c}{Model 5} \\ \hline
\multicolumn{3}{c}{ } \\[-2.0ex]
& 1000 & & & 11,319 & & & 10,884 & & & 10,704 & & & 10,713 & & & 10,997
& & & 10,394 & \\
& 2000 & & & 25,396 & & & 24,134 & & & 23,590 & & & 23,564 & & & 24,376
& & & 23,171 & \\
& 5000 & & & 72,630 & & & 68,171 & & & 66,192 & & & 66,232 & & & 69,039
& & & 66,027 & \\
& 10000 & & & 159,105 & & & 148,211 & & & 143,305 & & & 143,578 & & &
149,187 & & & 143,165 &
\end{tabular}
\end{center}
}

Models 2 and 3 show smaller values of averages compared to model 1.
The dependence of the average on \textit{n} is known to be
\begin{equation}
 \overline{C}_n \ = \ a\,n \log_e n \ + \ \mbox{lower order},
\end{equation}
where $a$ is 2 for model 1 (Knuth \cite{knuth3}), $12/7$ for model 2
(Sedgewick and Flajolet \cite{sedfla}), and $12600/8027$ for model 3
(Durand \cite{durand}). So the reductions relative to model~1 are
expected to be asymptotically 14.3\% for model 2 and 21.5\% for
model 3. Our values of 6.8\% and 9.9\%, respectively, at \textit{n} =
10,000 indicate that we are yet far below the asymptotic region where
the $n \log n$ term dominates.

In the range up to \textit{n} = 10,000, model 4, where the sample size
for selection of the partitioning element increases stepwise with
\textit{n}, does not show a further reduction of average values.
We have to consider that the ratio of comparisons needed for the
selection of the partitioning element, relative to the cost of
partitioning, is between (asymptotically)
$1/(2q_{\mbox{\scriptsize min}})$ and $3/(2q_{\mbox{\scriptsize min}})$.
So for $q_{\mbox{\scriptsize min}}=5$, the total cost (selection plus
partitioning) will be between 10\% and 30\% above that for the
partitioning alone. As our data indicate, this additional cost is largely
compensated by the improved balancing of the recursion tree.

In the range between \textit{n} = 1000 and 10,000, the data for model 5
show a reduction relative to model~4b---due to the usage of insertion
sort for small arrays---that is decreasing from 5.5\% to 4.0\%.
The decrease results from the fact that the share of work
done in the leaf nodes of the recursion tree is decreasing with
increasing \textit{n}.

\subsection*{\textnormal{\textsl{6.3 \ Standard deviation}}}

To determine standard deviations for the numbers of comparisons, defined
as
\begin{equation}
 \sigma_n \ \stackrel{\mathrm{def}}{=} \ \sqrt{\frac{1}{n!}
 \sum_{j \ge 0} (j-\overline{C}_n)^2 f_n(j)} ,
\end{equation}
we first had to compute frequency distributions $f_n$ from the recurrence
equation (\ref{rec1}) (plus recurrence (\ref{rec4}) in the case of
model 5), which was done for values of \textit{n} up to 500. Results are
shown in Figure~6. For models 1 to 3, the dependence on \textit{n}
appears almost linear. For a simple version of Quicksort corresponding
to our model~1, Iliopoulos and Penman \cite{iliop} proved the formula
\begin{equation}
 \sigma_n \ = \ \sqrt{7n^2 - 4(n+1)^2 H_n^{(2)} - 2(n+1)H_n + 13n} ,
\end{equation}
which is compatible with our data (at \textit{n} = 500, the difference
is less than 0.1\%). For models 2 and 3, we find values that are lower
by factors of approximately $1/2$ and $1/4$, respectively. So not only
the maximum numbers of comparisons (Figure 3), but also the standard
deviations appear to decrease by a factor $1/2$ if the sample size is
tripled.

For models 4 and 5, the increase with \textit{n} is less than linear as
could be expected because the sample size is stepwise increasing with
\textit{n}. The steps at $3^k q_{\mbox{\scriptsize min}}$ are visible in
the figure, too. Due to the fact that the frequency distribution of
comparisons is broader for insertion sort than that of the---in the range
\mbox{$n \le n^{\mbox{\scriptsize b}}_{\mbox{\scriptsize max}}$}---
competing Quicksort with median of three selection used in model 4b,
model~5 shows greater values of the standard deviation. For a suitably
chosen value of
$n^{\mbox{\scriptsize b}}_{\mbox{\scriptsize max}}$, however, the
\linebreak average number of comparisons for selection sort is smaller
in this range of \textit{n}, so model 5 remains an improvement in
efficiency.

\subsection*{\textnormal{\textsl{6.4 \ Probability of bad cases}}}

Probabilities of bad cases were determined from frequency distributions
of comparisons. In Figure 7, we show some of these distributions
at \textit{n} = 500. The figure illustrates both the stronger than
exponential decrease in the tails of the distributions and the increase
of concentration as result of a greater effort to select the partitioning
element. Incidentally, for model 1 there is exactly one worst case, which
arises if always the element of largest value is picked for partitioning.
This leads to $(n$$+$$2)(n$$-$$1)/2$ comparisons.

\smallskip

What is a bad case of input data? An attempt to define such a case may
be, for example, that it should be one that needs more than 1.5 times the
average number of comparisons. This would, however, appear rather
artifical, because for some applications even smaller deviations might be
unacceptable whereas for others a greater factor might be tolerable.
For this reason we introduce a parameter, the ``threshold factor''
$\tau$, to discriminate between bad cases and acceptable ones.

\smallskip

As ``threshold factor'' or ``relative threshold value'' we define
\begin{equation}
 \tau \ \stackrel{\mathrm{def}}{=} \ C^{\mbox{\scriptsize thr}}
 \ / \  \overline{C}_n ,
\end{equation}
where $C^{\mbox{\scriptsize thr}}$ is the absolute threshold value and
$\overline{C}_n$ the average number of comparisons. In Figure 8, we
display results for the probability of cases needing more comparisons
than those given by a value of $\tau$, in dependence on \textit{n}. So
the probability displayed is
\begin{equation}
 p_{n, \tau} \ \stackrel{\mathrm{def}}{=} \ \frac{1}{n!} \sum_{j >
 \tau \overline{C}_n} \! f_n(j) ,
\end{equation}
where $f_n$ is the frequency distribution defined in Section 3.

\begin{figure}[tb]
{\parindent0em \setlength{\unitlength}{1.0cm} \begin{picture}(14.34,7.65)
\put(2.92,1){\framebox(8.5,6){}}
\small \put(2.92,0.84){\makebox(0,0)[t]{0}}
\put(3.574,1){\line(0,1){0.107}}
\put(4.228,1){\line(0,1){0.107}}
\put(4.882,1){\line(0,1){0.164}}
\put(4.882,0.84){\makebox(0,0)[t]{30,000}}
\put(5.535,1){\line(0,1){0.107}}
\put(6.189,1){\line(0,1){0.107}}
\put(6.843,1){\line(0,1){0.164}}
\put(6.843,0.84){\makebox(0,0)[t]{60,000}}
\put(7.497,1){\line(0,1){0.107}}
\put(8.151,1){\line(0,1){0.107}}
\put(8.805,1){\line(0,1){0.164}}
\put(8.805,0.84){\makebox(0,0)[t]{90,000}}
\put(9.458,1){\line(0,1){0.107}}
\put(10.112,1){\line(0,1){0.107}}
\put(10.766,1){\line(0,1){0.164}}
\put(10.766,0.84){\makebox(0,0)[t]{120,000}}
\put(3.574,7){\line(0,-1){0.107}}
\put(4.228,7){\line(0,-1){0.107}}
\put(4.882,7){\line(0,-1){0.164}}
\put(5.535,7){\line(0,-1){0.107}}
\put(6.189,7){\line(0,-1){0.107}}
\put(6.843,7){\line(0,-1){0.164}}
\put(7.497,7){\line(0,-1){0.107}}
\put(8.151,7){\line(0,-1){0.107}}
\put(8.805,7){\line(0,-1){0.164}}
\put(9.458,7){\line(0,-1){0.107}}
\put(10.112,7){\line(0,-1){0.107}}
\put(10.766,7){\line(0,-1){0.164}}
\put(2.76,1){\makebox(0,0)[r]{\xpten{0}}}
\put(2.92,1.5){\line(1,0){0.107}}
\put(2.92,2){\line(1,0){0.164}}
\put(2.76,2){\makebox(0,0)[r]{\xpten{200}}}
\put(2.92,2.5){\line(1,0){0.107}}
\put(2.92,3){\line(1,0){0.164}}
\put(2.76,3){\makebox(0,0)[r]{\xpten{400}}}
\put(2.92,3.5){\line(1,0){0.107}}
\put(2.92,4){\line(1,0){0.164}}
\put(2.76,4){\makebox(0,0)[r]{\xpten{600}}}
\put(2.92,4.5){\line(1,0){0.107}}
\put(2.92,5){\line(1,0){0.164}}
\put(2.76,5){\makebox(0,0)[r]{\xpten{800}}}
\put(2.92,5.5){\line(1,0){0.107}}
\put(2.92,6){\line(1,0){0.164}}
\put(2.76,6){\makebox(0,0)[r]{\xpten{1000}}}
\put(2.92,6.5){\line(1,0){0.107}}
\put(2.76,7){\makebox(0,0)[r]{\xpten{1200}}}
\put(11.42,1.5){\line(-1,0){0.107}}
\put(11.42,2){\line(-1,0){0.164}}
\put(11.42,2.5){\line(-1,0){0.107}}
\put(11.42,3){\line(-1,0){0.164}}
\put(11.42,3.5){\line(-1,0){0.107}}
\put(11.42,4){\line(-1,0){0.164}}
\put(11.42,4.5){\line(-1,0){0.107}}
\put(11.42,5){\line(-1,0){0.164}}
\put(11.42,5.5){\line(-1,0){0.107}}
\put(11.42,6){\line(-1,0){0.164}}
\put(11.42,6.5){\line(-1,0){0.107}}
\qbezier(3.149,5.363)(3.159,5.929)(3.168,6.495)
\qbezier(3.168,6.495)(3.178,6.56)(3.188,6.625)
\qbezier(3.188,6.625)(3.212,6.64)(3.236,6.655)
\qbezier(3.236,6.655)(3.683,6.574)(4.13,6.44)
\qbezier(4.13,6.44)(4.534,6.318)(5.11,6.095)
\qbezier(5.11,6.095)(5.635,5.892)(6.091,5.691)
\qbezier(6.091,5.691)(6.582,5.475)(7.072,5.229)
\qbezier(7.072,5.229)(7.578,4.974)(8.053,4.702)
\qbezier(8.053,4.702)(8.597,4.389)(9.033,4.089)
\qbezier(9.033,4.089)(9.542,3.738)(10.014,3.331)
\qbezier(10.014,3.331)(10.423,2.979)(10.668,2.643)
\qbezier(10.668,2.643)(10.762,2.513)(10.897,2.301)
\qbezier(10.897,2.301)(10.983,2.165)(11.028,2.025)
\qbezier(11.028,2.025)(11.06,1.921)(11.093,1.765)
\qbezier(11.093,1.765)(11.1,1.383)(11.106,1)
\small \put(9.422,4){\makebox(0,0)[l]{$1$}}
\qbezier(3.186,5.676)(3.203,6.163)(3.219,6.65)
\qbezier(3.219,6.65)(3.225,6.655)(3.232,6.66)
\qbezier(3.232,6.66)(3.237,6.657)(3.242,6.655)
\qbezier(3.242,6.655)(3.279,6.645)(3.317,6.635)
\qbezier(3.317,6.635)(3.448,6.577)(3.578,6.52)
\qbezier(3.578,6.52)(3.775,6.405)(3.971,6.29)
\qbezier(3.971,6.29)(4.167,6.155)(4.363,6.02)
\qbezier(4.363,6.02)(4.559,5.865)(4.755,5.711)
\qbezier(4.755,5.711)(4.952,5.539)(5.148,5.367)
\qbezier(5.148,5.367)(5.344,5.175)(5.54,4.984)
\qbezier(5.54,4.984)(5.736,4.767)(5.932,4.55)
\qbezier(5.932,4.55)(6.128,4.298)(6.325,4.045)
\qbezier(6.325,4.045)(6.521,3.727)(6.717,3.409)
\qbezier(6.717,3.409)(6.754,3.336)(6.791,3.262)
\qbezier(6.791,3.262)(6.85,3.125)(6.908,2.988)
\qbezier(6.908,2.988)(6.948,2.872)(6.987,2.756)
\qbezier(6.987,2.756)(7.007,2.679)(7.026,2.603)
\qbezier(7.026,2.603)(7.046,2.468)(7.065,2.332)
\qbezier(7.065,2.332)(7.069,2.151)(7.072,1.969)
\small \put(6.489,4){\makebox(0,0)[l]{$2$}}
\qbezier(3.194,5.637)(3.197,5.999)(3.2,6.215)
\qbezier(3.2,6.215)(3.207,6.635)(3.22,6.645)
\qbezier(3.22,6.645)(3.226,6.652)(3.232,6.66)
\qbezier(3.232,6.66)(3.24,6.669)(3.259,6.65)
\qbezier(3.259,6.65)(3.263,6.646)(3.39,6.55)
\qbezier(3.39,6.55)(3.463,6.495)(3.586,6.355)
\qbezier(3.586,6.355)(3.694,6.232)(3.782,6.12)
\qbezier(3.782,6.12)(3.88,5.995)(3.978,5.854)
\qbezier(3.978,5.854)(4.077,5.71)(4.174,5.553)
\qbezier(4.174,5.553)(4.272,5.393)(4.37,5.215)
\qbezier(4.37,5.215)(4.481,5.014)(4.566,4.833)
\qbezier(4.566,4.833)(4.642,4.671)(4.763,4.385)
\qbezier(4.763,4.385)(4.891,4.078)(4.959,3.82)
\qbezier(4.959,3.82)(5.016,3.601)(5.024,3.576)
\qbezier(5.024,3.576)(5.072,3.429)(5.089,3.255)
\qbezier(5.089,3.255)(5.112,3.037)(5.134,2.577)
\small \put(5.019,4){\makebox(0,0)[l]{$3$}}
\qbezier(3.156,5.051)(3.172,6.098)(3.188,6.445)
\qbezier(3.188,6.445)(3.2,6.687)(3.223,6.66)
\qbezier(3.223,6.66)(3.238,6.643)(3.254,6.625)
\qbezier(3.254,6.625)(3.288,6.583)(3.352,6.45)
\qbezier(3.352,6.45)(3.414,6.321)(3.45,6.235)
\qbezier(3.45,6.235)(3.492,6.133)(3.548,5.975)
\qbezier(3.548,5.975)(3.601,5.824)(3.646,5.682)
\qbezier(3.646,5.682)(3.691,5.539)(3.744,5.351)
\qbezier(3.744,5.351)(3.804,5.137)(3.842,4.971)
\qbezier(3.842,4.971)(3.891,4.749)(3.94,4.527)
\qbezier(3.94,4.527)(4.012,4.242)(4.038,3.976)
\qbezier(4.038,3.976)(4.072,3.627)(4.107,2.923)
\small \put(4.135,4){\makebox(0,0)[l]{$5$}}
\normalsize \put(2.92,7.36){\makebox(0,0)[bl]{Frequency distributions of comparisons for $n=500$}}
\small \put(7.17,0.4){\makebox(0,0)[t]{$c$ : Number of comparisons}}
\put(1.56,4){\makebox(0,0)[r]{$f_{500}(c)$}}
\end{picture}
}
{\small Figure 7. Frequency distributions of comparisons for \textit{n}
 = 500. Curve labels are model numbers \\
($q_{\mbox{\scriptsize min}} = 5$ for models 3 and 5,
$n^{\mbox{\scriptsize b}}_{\mbox{\scriptsize max}} = 9$ for model 5).}

\vspace{1.7cm}

\input{figure8}
{\small Figure 8. Relative numbers of cases needing more comparisons
than the number given by a threshold factor $\tau$, in dependence on
\textit{n}. Curve labels indicate the threshold factor.}
\end{figure}

\clearpage

\begin{figure}[b]
\input{figure9}
{\small Figure 9. Relative numbers of cases needing more comparisons
than the number given by a threshold factor for \textit{n}~=~100, 250,
and 500, in dependence on the threshold factor.}
\end{figure}

\smallskip

For models 1, 2, 3, and 5, the figure shows function graphs of
$p_{n, \tau}$ in dependence on \textit{n}, for thresholds that are 10\%,
25\%, and 50\% above the average number of comparisons. As a counterpart
to this, Figure~9 displays probabilities for some values of \textit{n}
in dependence on~$\tau$. These two figures contain---for some given
models, values of \textit{n} and values of~$\tau$---an answer to the
question posed in the title of this article.

\smallskip

The sequence of Figures 8a to 8d reveals the following facts:
\begin{enumerate}
\item  For greater values of the threshold factor,
the decrease of $p_{n, \tau}$ in dependence on \textit{n} is
stronger than that for smaller ones. \\[-1.6em]
\item  A greater sample size for the selection of the partitioning
element leads to smaller probability values for bad cases and to a
stronger decrease in \textit{n}. \\[-1.6em]
\item  The adaptive method of model 5 leads to step effects at numbers
of elements $n = 3^k q_{\mbox{\scriptsize min}}$.
\end{enumerate}

As a supplement to Figure 8 we present the following table which shows
values for the ratio $p_{500, \tau} / p_{250, \tau}$:
{
\begin{center}
\begin{tabular}{clcclcclcclcclc}
\multicolumn{3}{c}{$\tau$} &
\multicolumn{3}{c}{Model 1} &
\multicolumn{3}{c}{Model 2} &
\multicolumn{3}{c}{Model 3} &
\multicolumn{3}{c}{Model 5} \\ \hline
\multicolumn{3}{c}{ } \\[-2.0ex]
 & 1.1 & & & 0.78 & & & 0.64 & & & 0.38 & & & \xnum{4.44}{-4} & \\
 & 1.25 & & & 0.45 & & & 0.26 & & & 0.058 & & & \xnum{1.60}{-10} & \\
 & 1.5 & & & 0.15 & & & 0.045 & & & \xnum{1.89}{-3} & & &
 \xnum{2.16}{-22} & \\
 & 2.0 & & & 0.012 & & & \xnum{9.24}{-4} & & & \xnum{2.63}{-6} & & &
 \xnum{1.90}{-51} &
\end{tabular}
\end{center}
}
The table, too, indicates that the decrease of the relative number of bad
cases with increasing \textit{n} is strongly influenced by the value of
$\tau$ and by the selection method.

\medskip

Figure 9 shows the probability of bad cases for \textit{n} = 100, 250,
and 500 in dependence on the threshold factor for values of $\tau$ up
to 3. The decrease of the probabilities with increasing $\tau$ turns
out to be stronger than exponential as could be expected from the
frequency distributions of comparisons (Figure 7). Even more than
Figure 8, Figure 9 emphasizes the enormous differences in the order of
magnitude among the different versions.

\medskip

Additionally to these two figures, we list some probability values for
bad cases at \textit{n} = 500 in the following table:

{
\begin{center}
\begin{tabular}{clcclcclcclcclc}
\multicolumn{3}{c}{$\tau$} &
\multicolumn{3}{c}{Model 1} &
\multicolumn{3}{c}{Model 2} &
\multicolumn{3}{c}{Model 3} &
\multicolumn{3}{c}{Model 5} \\ \hline
\multicolumn{3}{c}{ } \\[-2.0ex]
 & 1.1 & & & \xnum{7.35}{-2} & & & \xnum{1.14}{-2}
 & & & \xnum{2.83}{-4} & & & \xnum{3.75}{-8} & \\
 & 1.25 & & & \xnum{2.17}{-3} & & & \xnum{8.88}{-6}
 & & & \xnum{1.28}{-10} & & & \xnum{2.97}{-23} & \\
 & 1.5 & & & \xnum{1.88}{-6} & & & \xnum{5.10}{-12}
 & & & \xnum{2.17}{-23} & & & \xnum{5.12}{-53} & \\
 & 2.0 & & & \xnum{1.02}{-13} & & & \xnum{6.66}{-27}
 & & & \xnum{3.62}{-54} & & & \xnum{4.25}{-127} &
\end{tabular}
\end{center}
}

We try to support the judgement on these probability values in the
following way. Assume that at intervals of one millisecond sorts of
random data are started. Then we ask for the expected time until one
event exceeding a given relative threshold value will occur. To give an
example: for model 2 and \textit{n}~=~500, the expected time to
encounter one case needing more comparisons than 1.25 times the average
is 1.9 minutes. The following table presents such expected times for
\textit{n} = 500 (s: seconds, m: minutes, h: hours, d: days, a: years):

{
\begin{center}
\begin{tabular}{clcclcclcclcclc}
\multicolumn{3}{c}{$\tau$} &
\multicolumn{3}{c}{Model 1} &
\multicolumn{3}{c}{Model 2} &
\multicolumn{3}{c}{Model 3} &
\multicolumn{3}{c}{Model 5} \\ \hline
\multicolumn{3}{c}{ } \\[-2.0ex]
 & 1.1 & & & 0.014 s & & & 0.09 s & & & 3.5 s & & & 7.4 h & \\
 & 1.25 & & & 0.46 s & & & 1.9 m & & & 90.5 d & & & \xnum{1.1}{12} a & \\
 & 1.5 & & & 8.9 m & & & 6.2 a & & & \xnum{1.5}{12} a & & &
 \xnum{6.2}{41} a & \\
 & 2.0 & & & 312 a & & & \xnum{4.8}{15} a & & & \xnum{8.8}{42} a & & &
 \xnum{7.5}{115} a &
\end{tabular}
\end{center}
}

Here it becomes clear that values for model 2, $\tau \ge 2$, model 3,
$\tau \ge 1.5$, and model 5, $\tau \ge 1.25$ are beyond every
imagination. For comparison: the age of our universe is ``only''
\xnum{1.4}{10} years! And these are results for the rather small number
\textit{n} = 500. Probabilities will become even smaller if \textit{n}
is increased to values which are worth using the Quicksort algorithm for
sorting.

\smallskip

It should be kept in mind, however, that these are results for uniformly
distributed random data of distinct values. In fact, data that are not
pairwise distinct are not a real problem if an adequate method for
partitioning is used, as will be shown in Section 7. The main pitfall is
the fact that data showing up in practice tend to possess some order
pattern that might lead to a bad case. An attempt to solve this problem
has been randomization as shown already in Hoare's first articles on
Quicksort \cite{hoare61}. On the other hand, this does not change the
statistical probability for bad cases. Our adaptive recursive median of
medians approach not only reduces the probabilities for such cases
statistically to values that are virtually zero but also works against
non-random data being preferred candidates for them.

\begin{figure}[t]
\vspace*{0.5cm}
{\small
\begin{verbatim}
   /* ipart: index of the partitioning element */
      partval = a[ipart];
      a[ipart] = a[0];

      i = 0;
      for (j = 1; j < n; j++)
         if (a[j] < partval) {
            temp = a[++i]; a[i] = a[j]; a[j] = temp;
         }
      a[0] = a[i];
      a[i] = partval;

   /* subarray 1: a[0..(i-1)],  subarray 2: a[(i+1)..(n-1)] */
\end{verbatim}
\vspace{-5.56cm}
{\parindent0em \setlength{\unitlength}{1.0cm} \begin{picture}(14.31,5.65)
\put(0,0){\framebox(14.31,5.65){}}
\end{picture}
}}
\begin{center} {\small Figure 10. Partitioning algorithm (simple sweep
method)} \end{center}

 \vspace{0.23in}

\noindent
{\small
\begin{verbatim}
      partval = a[ipart];

      j = 0;
      for (i = 0; i < n; i++)
         if (a[i] < partval) {
            temp = a[j]; a[j++] = a[i]; a[i] = temp;
         }

      i = j;
      for (k = i; k < n; k++)
         if (a[k] == partval) {
            temp = a[i]; a[i++] = a[k]; a[k] = temp;
         }

   /* subarray 1: a[0..(j-1)],  subarray 2: a[i..(n-1)] */
\end{verbatim}
\vspace{-6.23cm}
{\parindent0em \setlength{\unitlength}{1.0cm} \begin{picture}(14.31,6.40)
\put(0,0){\framebox(14.31,6.40){}}
\end{picture}
}}
\begin{center} {\small Figure 11. Partitioning algorithm (extended sweep
method -- three-way)} \end{center}
 \vspace{-0.1in}
\end{figure}

\section*{\textnormal{7. \ Characteristics of partitioning algorithms}}

Partitioning algorithms may be divided into two categories, which we
call ``sweep methods'' and ``collision methods'', respectively, to have
short expressive names at hand. With sweep methods, a scan is done by
``sweeping'' a pointer or index over the array from left to right.
Collision methods use two pointers or indices, one for scanning from
left to right, one from right to left, until they collide.

Simple sweep methods are shown in some textbooks, e.\,g.
\cite{KR, cormen}. Bentley \cite{bent99} tells us that he learned such a
scheme from Nico Lomuto. The algorithm displayed in Figure~10 produces
the arrangement
\begin{displaymath}
 \underbrace{\mbox{elements} < \mbox{p,}}_{\mbox{subarray 1}} \
 \mbox{p,} \ \underbrace{\mbox{elements} \ge
 \mbox{p}}_{\mbox{subarray 2}}.
\end{displaymath}
So elements equal to the partitioning element p (except p itself)
will be placed into subarray 2. If all values of elements are equal,
Quicksort needs the maximally possible number of $n(n$$-$$1)/2$
comparisons with this method. Since our objective is a general purpose
version for sorting, a poor performance for ``familiar'' non-random
cases is unacceptable: sequences which are, for example, (nearly)
sorted (ascending or descending), have an organ pipe characteristic
(as mentioned by Bentley and McIlroy \cite{bent93}), or all equal
values, should be processed at speeds not slower than the average.
Hence we cannot consider this method as suitable for use in practice.

In an attempt to get a practically usable version, the algorithm may be
extended to perform a ``three-way partitioning'' so that the arrangement
\begin{equation}
 \underbrace{\mbox{elements} < \mbox{p,}}_{\mbox{subarray 1}} \
 \mbox{elements = p,} \ \underbrace{\mbox{elements} >
 \mbox{p}}_{\mbox{subarray 2}}
\end{equation}
is achieved. This can be done by two loops in sequence as shown in
Figure 11.

Sweep methods are optimal with respect to comparisons. The algorithm
of Figure 10 always needs $n$$-$$1$ comparisons, that of Figure 11
needs $(3n$$+$$1)/2$ comparisons on average (see also Tables 1 and 2
on page \pageref{tables12} for counting results). However, they have the
disadvantage that elements may be moved several times until they reach
their final position. Collision methods avoid this waste of resources:
at the cost of a few additional comparisons they reduce the number of
data movements---defined as assignments in which one or both operands
are array elements---asymptotically by a factor of three.

Already in his first series of papers on Quicksort, Hoare \cite{hoare61}
published a collision algorithm for partitioning. In the following years,
attempts were made to improve this method (Hibbard \cite{hibb63},
Scowen \cite{scowen}, Singleton \cite{singl}, van Emden \cite{emden}),
of which Singleton's version has become standard and was reproduced in
various textbooks (e.\,g. Wirth \cite{wirth75, wirth86}, Sedgewick
\cite{sedg90}, Cormen et al. \cite{cormen}). Sometimes it is
inaccurately attributed to Hoare. A listing of this algorithm, which we
call the ``classic collision method'', is shown in Figure 12.

It turns out that for arrays of \textit{n} pairwise different elements,
arrangement (4), shown in Section 3, is achieved with this method if
and only if the partitioning element happens to be in its final position
already from the beginning. In this case \textit{n}+1 comparisons are
needed. In all other cases arrangement (2), shown in Section 2.1, will
result, and the number of comparisons is either \textit{n} or
\textit{n}+2. For example, counting over all permutations of
\textit{n} = 10 elements (partitioning elements taken from index
$\lfloor n/2\rfloor$) yields the distribution

\smallskip

{
\tabcolsep0.15in
\begin{center}
\begin{tabular}{lccc}
Comparisons: & 10 & 11 & 12 \\
Frequency: & 756,000 & 362,880 & 2,509,920
\end{tabular}
\end{center}
}

\noindent where the \textit{n}+1 = 11 comparisons occur in the
$(n$$-$$1)!$ cases in which the partitioning element is in its
final position already before the partitioning process. In order to
always achieve arrangement (1) (Section~2.1), Scowen \cite{scowen}
suggested to swap the partitioning element, which he took from the
middle of the array, to one end and swap it to its correct position
afterwards. Sedgewick \cite{sedg78, sedg98} essentially combined this
with Singleton's version. However, in his simplest version of Quicksort
he used the element already at the end of the array for partitioning,
which leads to a worst case behavior for sorted sequences (ascending or
descending) and gave rise to the rumor that this is a general property
of Quicksort.

Our version (Figure 1), which we now call ``new collision method'',
has been derived from Sedge\-wick's algorithm. For randomly chosen
partitioning elements, all final index positions have equal probability,
so equations (3) lead to the average number of $n$$-$$1/2$ comparisons.
Sedgewick's version needs $n$$+$$1$$-$$1/n$ comparisons on average.

Since the algorithm always produces arrangement (1), it is preferable
to the classic collision method. Table 1 shows counting results for this
version, too.

These algorithms can also be transformed into three-way versions by
applying the two-pass strategy that was already shown for the sweep
method. Figure 13 displays such a version for the classic collision
algorithm. In their implementation of the C library function
\verb+qsort+, Bentley and McIlroy \cite{bent93} used a more complicated
technique, which needs more resources as can be seen in Table 2. In
contrast to the extended classic collision algorithm, it may move
elements back and forth even if the three-way arrangement (47) already
exists. For \verb+qsort+, however, it is the method of choice because it
benefits from the three-way comparison results (less, greater, equal)
which the user supplied comparison function has to deliver.
Incidentally, it should be noted that the popular method to produce
a three-way comparison result for integer operands by taking the
difference may lead to arithmetic overflows for certain pairs of operand
values; so it is not applicable within a library program.

\begin{figure}[t]
\vspace*{0.5cm}
{\small
\begin{verbatim}
      partval = a[ipart];
      i = 0;
      j = n-1;
      do {
         while (a[i] < partval)
            i++;
         while (partval < a[j])
            j--;
         if (i <= j) {
            temp = a[i]; a[i++] = a[j]; a[j--] = temp;
         }
      } while (i <= j);

   /* subarray 1: a[0..j],  subarray 2: a[i..(n-1)] */
\end{verbatim}
\vspace{-5.88cm}
{\parindent0em \setlength{\unitlength}{1.0cm} \begin{picture}(14.31,5.97)
\put(0,0){\framebox(14.31,5.97){}}
\end{picture}
}}
\begin{center} {\small Figure 12. \label{fig12} Partitioning algorithm
(classic collision method, Singleton \cite{singl})} \end{center}
\end{figure}

\begin{figure}[b]
{\small
\begin{verbatim}
      partval = a[ipart];
      j = 0;
      k = n-1;
      done = FALSE;
      do {
         while (a[j] < partval)
            j++;
         while (j < k && partval <= a[k])
            k--;
         if (k <= j)
            done = TRUE;
         else {
            temp = a[j]; a[j++] = a[k]; a[k--] = temp;
         }
      } while (!done);

      k = j;
      i = n-1;
      done = FALSE;
      do {
         while (k <= i && partval == a[k])
            k++;
         while (k <= i && partval < a[i])
            i--;
         if (i <= k)
            done = TRUE;
         else {
            temp = a[k]; a[k++] = a[i]; a[i--] = temp;
         }
      } while (!done);

   /* subarray 1: a[0..(j-1)],  subarray 2: a[(i+1)..(n-1)] */
\end{verbatim}
\vspace{-12.55cm}
{\parindent0em \setlength{\unitlength}{1.0cm} \begin{picture}(14.31,12.65)
\put(0,0){\framebox(14.31,12.65){}}
\end{picture}}
}
\begin{center} {\small Figure 13. Partitioning algorithm (extended
collision method -- three-way)} \end{center}
\end{figure}

\clearpage

\begin{figure}[t]
{\tabcolsep0.2in
\begin{center}
\begin{tabular}{|r|r|r|r|r|r|r|} \hline
  & \multicolumn{2}{|c|}{\rule[-2mm]{0mm}{6mm}Sweep simple}
  & \multicolumn{2}{|c|}{Classic collision}
  & \multicolumn{2}{|c|}{Collision new} \\[0.5ex] \hline
 \rule[-2mm]{0mm}{6mm}\textit{n} & $C_n^{\mbox{\scriptsize avg}}$
  & $M_n^{\mbox{\scriptsize avg}}$
  & $C_n^{\mbox{\scriptsize avg}}$ & $M_n^{\mbox{\scriptsize avg}}$
  & $C_n^{\mbox{\scriptsize avg}}$ & $M_n^{\mbox{\scriptsize avg}}$
  \\[0.5ex] \hline & & & & & & \\[-2.2ex]
    2  &  1.0 &  5.5  &    2.500 & 4.000  &    1.5 &  5.0 \\
    3  &  2.0 &  7.0  &    3.333 & 4.500  &    2.5 &  5.5 \\
    4  &  3.0 &  8.5  &    4.917 & 4.750  &    3.5 &  6.0 \\
    5  &  4.0 & 10.0  &    6.200 & 5.200  &    4.5 &  6.5 \\
    6  &  5.0 & 11.5  &    7.300 & 5.600  &    5.5 &  7.0 \\
    7  &  6.0 & 13.0  &    8.381 & 6.071  &    6.5 &  7.5 \\
    8  &  7.0 & 14.5  &    9.423 & 6.518  &    7.5 &  8.0 \\
    9  &  8.0 & 16.0  &   10.460 & 7.000  &    8.5 &  8.5 \\
   10  &  9.0 & 17.5  &   11.483 & 7.467  &    9.5 &  9.0 \\
   11  & 10.0 & 19.0  &   12.505 & 7.955  &   10.5 &  9.5 \\
   12  & 11.0 & 20.5  &   13.520 & 8.432  &   11.5 & 10.0 \\
   13  & 12.0 & 22.0  &   14.534 & 8.923  &   12.5 & 10.5 \\
   14  & 13.0 & 23.5  &   15.544 & 9.407  &   13.5 & 11.0 \\
   15  & 14.0 & 25.0  &   16.554 & 9.900  &   14.5 & 11.5 \\ \hline
\end{tabular}
\end{center}
}
 \label{tables12}
{\small Table 1. Average numbers of comparisons
$C_n^{\mbox{\scriptsize avg}}$ and data movements
$M_n^{\mbox{\scriptsize avg}}$, needed by three different partitioning
algorithms. Values have been obtained by counting over all permutations
of \textit{n} pairwise different elements. Partitioning elements are
always taken from index $\lfloor n/2\rfloor$). The algorithms are listed
in Figures 10, 12, and 1, respectively.}
\end{figure}

\begin{figure}[b]
{\tabcolsep0.2in
\begin{center}
\begin{tabular}{|r|r|r|r|r|r|r|} \hline
  & \multicolumn{2}{|c|}{\rule[-2mm]{0mm}{6mm}Sweep extended}
  & \multicolumn{2}{|c|}{Classic collision ext.}
  & \multicolumn{2}{|c|}{Bentley \& McIlroy} \\[0.5ex]
  \hline
 \rule[-2mm]{0mm}{6mm}\textit{n} & $C_n^{\mbox{\scriptsize avg}}$
  & $M_n^{\mbox{\scriptsize avg}}$
  & $C_n^{\mbox{\scriptsize avg}}$ & $M_n^{\mbox{\scriptsize avg}}$
  & $C_n^{\mbox{\scriptsize avg}}$ & $M_n^{\mbox{\scriptsize avg}}$
  \\[0.5ex] \hline & & & & & & \\[-2.2ex]
    2  &  3.5 &  5.5  &   3.500 &  2.500  &   4.500 &  7.0 \\
    3  &  5.0 &  7.0 &  5.667 &  3.500 &  6.167 &  7.5 \\
    4  &  6.5 &  8.5 &  7.333 &  4.250 &  7.833 &  8.0 \\
    5  &  8.0 & 10.0 &  8.950 &  4.900 &  9.500 &  8.5 \\
    6  &  9.5 & 11.5 & 10.533 &  5.500 & 11.167 &  9.0 \\
    7  & 11.0 & 13.0 & 12.057 &  6.143 & 12.833 &  9.5 \\
    8  & 12.5 & 14.5 & 13.613 &  6.679 & 14.500 & 10.0 \\
    9  & 14.0 & 16.0 & 15.111 &  7.321 & 16.167 & 10.5 \\
   10  & 15.5 & 17.5 & 16.653 &  7.825 & 17.833 & 11.0 \\
   11  & 17.0 & 19.0 & 18.144 &  8.464 & 19.500 & 11.5 \\
   12  & 18.5 & 20.5 & 19.676 &  8.950 & 21.167 & 12.0 \\
   13  & 20.0 & 22.0 & 21.167 &  9.581 & 22.833 & 12.5 \\
   14  & 21.5 & 23.5 & 22.693 & 10.058 & 24.500 & 13.0 \\
   15  & 23.0 & 25.0 & 24.186 & 10.681 & 26.167 & 13.5 \\ \hline
\end{tabular}
\end{center}
}
{\small Table 2. Average numbers of comparisons
$C_n^{\mbox{\scriptsize avg}}$ and data movements
$M_n^{\mbox{\scriptsize avg}}$, needed by three different three-way
partitioning algorithms. Values have been obtained by counting over all
permutations of \textit{n} pairwise different \linebreak elements.
Partitioning elements are always taken from index $\lfloor n/2\rfloor$).
Algorithms ``sweep extended'' and ``classic collision extended'' are
listed in Figures 11 and 13, respectively.}
\end{figure}

\clearpage

The following table shows the frequencies of comparisons obtained by
counting over all permutations of \textit{n} = 10 pairwise different
elements for two three-way algorithms (partitioning elements taken from
index $\lfloor n/2\rfloor$).
In both cases the distribution appears irregular:
{
\tabcolsep0.15in
\begin{center}
\begin{tabular}{|crc|crc|crc|} \hline
\multicolumn{3}{|c|}{\rule[-2mm]{0mm}{6mm}Comparisons} &
\multicolumn{3}{|c|}{Classic collision ext.} &
\multicolumn{3}{|c|}{Bentley \& McIlroy} \\ \hline
\multicolumn{3}{|c|}{ } & \multicolumn{3}{|c|}{ } &
\multicolumn{3}{|c|}{ } \\[-2.0ex]
& 12 & & & 403,200 & & &  25,920 & \\
& 13 & & & 141,120 & & &   2,880 & \\
& 14 & & & 413,280 & & & 285,120 & \\
& 15 & & & 252,000 & & & 118,080 & \\
& 16 & & & 806,400 & & & 622,080 & \\
& 17 & & & 161,280 & & & 535,680 & \\
& 18 & & & 241,920 & & & 518,400 & \\
& 19 & & & 403,200 & & & 679,680 & \\
& 20 & & & 403,200 & & & 362,880 & \\
& 21 & & & 403,200 & & & 478,080 & \\ \hline
\end{tabular}
\end{center}
}

\medskip

The first decision when choosing an algorithm should be that between
two-way and three-way partitioning. A three-way version will need, on
average, about 50\% more comparisons than a two-way method. In a
Quicksort implementation, this only pays off if the number of distinct
values of elements is small compared to \textit{n}. For our numerical
analysis we therefore decided for a two-way version.

Second, sweep methods should be dropped for practical use because of
their poor performance. From the remaining two-way algorithms that we
discussed, only the ``new collision'' method has
properties (equations (3)) to make it suitable for our analysis.
Because of its optimal partitioning---always arrangement (1) shown in
Section 2.1---it should also be favored in practice.

If the decision is for a three-way version and if comparisons of
elements by operators are done directly within the sorting program, we
suggest to use the ``extended classic collision'' method for efficiency
reasons. The advantage of Bentley and McIlroy's version for polymorphic
programs like \verb+qsort+ has already been pointed out
above.

\section*{\textnormal{8. \ Our Quicksort implementation}}

In the appendix, we present a Quicksort program conforming to the C99
standard \cite{ISO99} and compatible also with C++. It incorporates an
implementation of the recursive median of three medians method and is
meant to reduce the probability of bad cases to very small values which
can be regarded as virtually zero. In contrast to the C library function
\verb+qsort+, this version is not polymorphic, so the type of array
elements has to be fixed before compilation.

Parameterizations are used to support easy adaptation to the needs of
application programs. The type of array elements is declared in a single
\verb+typedef+ statement. Comparison operators for the element type are
defined as preprocessor macros.

The values of parameters $q_{\mbox{\scriptsize min}}$ (equation (33))
and $n^{\mbox{\scriptsize b}}_{\mbox{\scriptsize max}}$ (Section 4.5)
are set in \verb+enum+ statements for identifiers \verb+QMIN+ and
\verb+NBASISMAX+, respectively. Decreasing the value for \verb+QMIN+
will decrease the probability for bad cases, but will increase the
average sorting time. We recommend a value within the range 5 to 10.
Ideally, the optimum value for \verb+NBASISMAX+ would then be determined
by time measurements. In most cases, a value within the range 10 to 20
turns out to be reasonable.

The switch \verb+THREEWAY+ allows to choose between two-way and
three-way partitioning. The respective algorithms are those
shown in Figures 1 and 13.
The three-way version will, on average, need about 50\% more comparisons
for partitioning than its two-way counterpart. This is a waste of
resources if the ratio $n_d/n$, where $n_d$ is the number of distinct
values of elements, is near to one. So if this characteristic of data is
known for an application, the two-way version should be preferred,
whereas for a general purpose sorting function the three-way variant is
the method of choice.

For the median of three selection we used, apart from the element in the
middle of the array, those at index positions $\lfloor n/4\rfloor$ and
$\lfloor n/2\rfloor$$+$$\lfloor n/4\rfloor$ instead of the first and
last ones. It turned out that due to an interference with our
partitioning method, the usual version leads to a (nearly) worst case
behavior for decreasing element values, which is unacceptable for
our program.

Recursive calls of the sort function have been replaced by loops and
use of an explicit stack. This well-known technique was first
published by Hibbard \cite{hibb63}. However, determination of the
recursive median of medians is done by a recursive function, which
uses sample size \textit{m} = 9 as the only basis case. The recursion
depth for such a determination is
$\lfloor \log_3\,(n/q_{min})\rfloor$$-$$2$.

\subsection*{\textnormal{\textsl{Time measurements}}}

\renewcommand{\thefootnote}{\fnsymbol{footnote}}

Time measurements \label{reft3} for our implementation were carried out
on a desktop PC, installed in 2014 and running under Windows 7. We used
the C compiler from Microsoft Visual Studio
2012\footnote[2]{Surprisingly, the compiler does not support use of the
x86 extended precision data type that we need for our numerical
computations.},
always with optimization switch /O2. Installation parameters for
Quicksort were set---where applicable---as \verb+QMIN=5+ and
\verb+NBASISMAX=15+.

\begin{figure}[ht]
{
\tabcolsep0.07in
\begin{center}
\begin{tabular}{|r|rr|rr|rr|rr|rr|} \hline\hline
\multicolumn{1}{|c|}{\rule[0mm]{0mm}{4mm}} &
\multicolumn{2}{|c|}{Model 1} & \multicolumn{2}{|c|}{Model 2} &
\multicolumn{2}{|c|}{Model 3} & \multicolumn{2}{|c|}{Model 5} &
\multicolumn{2}{|c|}{Heapsort}\\
\multicolumn{1}{|c|}{\raisebox{1.5ex}[-1.5ex]{\textit{n}}} &
\multicolumn{1}{|c}{2-way} & \multicolumn{1}{c|}{3-way} &
\multicolumn{1}{|c}{2-way} & \multicolumn{1}{c|}{3-way} &
\multicolumn{1}{|c}{2-way} & \multicolumn{1}{c|}{3-way} &
\multicolumn{1}{|c}{2-way} & \multicolumn{1}{c|}{3-way} &
\multicolumn{1}{|c}{ \, classic} & \multicolumn{1}{c|}{BU}
\\[0.5ex] \hline\hline
\multicolumn{11}{|c|}{\rule[-2.5mm]{0mm}{7mm}{{\tt int} \ random}} \\
\hline  & & & & & & & & & & \\[-2.2ex]
100,000 & 5.6 & 6.7 \ & 5.9 & 6.8 \, & 6.0 & 7.1 \ & 6.2 & 6.9 \ &
5.5 & 7.3 \ \\
500,000 & 31.5 & 37.3 \ & 32.8 & 38.2 \ & 33.8 & 39.7 \ &
35.4 & 39.2 \ & 41.8 & 44.7 \ \\
1,000,000 & 66.0 & 77.9 \ & 69.0 & 79.6 \ & 70.7 & 82.8 \ &
75.1 & 83.0 \ & 101.1 & 97.7 \ \\
5,000,000 & 368.9 & 431.7 \ & 378.6 & 439.3 \ & 391.6 & 453.8 \ &
425.3 & 468.6 \ & 1121.8 & 802.6 \ \\ \hline
\multicolumn{11}{|c|}{\rule[-2.5mm]{0mm}{7mm}{{\tt int} \ increasing}}
\\ \hline
& & & & & & & & & & \\[-2.2ex]
100,000 & 0.9 & 1.4 \ & 1.0 & 1.5 \ & 1.0 & 1.5 \ & 0.9 & 1.4 \ &
4.2 & 4.4 \ \\
500,000 & 4.8 & 7.6 \ & 5.1 & 7.9 \ & 5.2 & 8.1 \ & 5.2 & 7.8 \ &
24.3 & 24.0 \ \\
1,000,000 & 10.1 & 15.9 \ & 10.6 & 16.4 \ & 10.9 & 16.9 \ &
11.1 & 16.6 \ & 51.0 & 49.4 \ \\
5,000,000 & 59.5 & 91.2 \ & 64.4 & 93.6 \ & 64.4 & 95.1 \ &
68.6 & 98.1 \ & 312.9 & 282.2 \ \\ \hline
\multicolumn{11}{|c|}{\rule[-2.5mm]{0mm}{7mm}{{\tt int} \ decreasing}}
\\ \hline
& & & & & & & & & & \\[-2.2ex]
100,000 & 1.0 & 1.5 \ & 1.1 & 1.5 \ & 1.1 & 1.5 \ & 1.0 & 1.4 \ &
4.0 & 4.8 \ \\
500,000 & 5.6 & 7.7 \ & 5.8 & 8.0 \ & 5.7 & 8.2 \ & 5.6 & 7.9 \ &
23.4 & 25.8 \ \\
1,000,000 & 11.6 & 16.1 \ & 12.1 & 16.7 \ & 11.9 & 17.2 \ &
11.8 & 16.8 \ & 51.9 & 53.9 \ \\
5,000,000 & 70.5 & 93.4 \ & 68.4 & 95.4 \ & 67.7 & 96.7 \ &
73.1 & 99.6 \ & 302.2 & 301.8 \ \\ \hline
\multicolumn{11}{|c|}{\rule[-2.5mm]{0mm}{7mm}{{\tt int} \ equal}}
\\ \hline
& & & & & & & & & & \\[-2.2ex]
100,000 & 1.5 & 0.1 \ & 1.5 & 0.1 \ & 1.5 & 0.1 \ & 1.3 & 0.1 \ &
0.3 & 2.4 \ \\
500,000 & 7.9 & 0.6 \ & 8.1 & 0.6 \ & 8.0 & 0.6 \ & 7.2 & 0.6 \ &
1.6 & 15.1 \ \\
1,000,000 & 16.6 & 1.1 \ & 17.0 & 1.1 \ & 16.7 & 1.1 \ &
15.2 & 1.3 \ & 3.3 & 38.0 \ \\
5,000,000 & 92.7 & 5.7 \ & 94.8 & 5.9 \ & 94.5 & 5.9 \ &
98.0 & 8.1 \ & 16.9 & 255.3 \ \\ \hline
\multicolumn{11}{|c|}{\rule[-2.5mm]{0mm}{7mm}{{\tt int} \ organ pipe}}
\\ \hline
& & & & & & & & & & \\[-2.2ex]
100,000 & 20.4 & 44.6 \ & 2.1 & 2.3 \ & 2.6 & 2.9 \ & 2.2 & 2.6 \ &
4.2 & 5.2 \ \\
500,000 & 803 & 3381 \ & 11.3 & 12.8 \ & 13.6 & 15.4 \ &
11.6 & 14.0 \ & 24.8 & 29.4 \ \\
1,000,000 & 3172 & 3560 \ & 23.7 & 26.3 \ & 27.9 & 31.8 \ &
23.9 & 29.0 \ & 53.3 & 61.4 \ \\
5,000,000 & 96539 & 69761 \ & 127.2 & 143.4 \ & 149.8 & 172.2 \ &
134.5 & 163.1 \ & 334.9 & 361.0 \ \\ \hline
\multicolumn{11}{|c|}{\rule[-2.5mm]{0mm}{7mm}{{\tt int} \ \{0,\,1\}
random}} \\ \hline
& & & & & & & & & & \\[-2.2ex]
100,000 & 1.8 & 0.4 \ & 1.9 & 0.5 \ & 1.9 & 0.4 \ & 1.8 & 0.5 \ &
2.5 & 3.5 \ \\
500,000 & 9.9 & 2.2 \ & 10.1 & 2.3 \ & 10.2 & 2.2 \ &
10.1 & 2.5 \ & 14.1 & 19.3 \ \\
1,000,000 & 20.5 & 4.5 \ & 21.0 & 4.5 \ & 21.0 & 4.5 \ &
21.2 & \ 5.2 \ & 29.3 & 41.6 \ \\
5,000,000 & 112.9 & 22.9 \ & 114.6 & 23.4 \ & 114.9 & 23.2 \ &
118.1 & \ 27.3 \ & 170.2 & 254.6 \ \\ \hline
\multicolumn{11}{|c|}{\rule[-2.5mm]{0mm}{7mm}{{\tt double} \ random}}
\\ \hline
& & & & & & & & & & \\[-2.2ex]
100,000 & 6.2 & 7.4 \ & 6.4 & 7.5 \ & 6.7 & 7.7 \ & 6.9 & 7.6 \ &
7.2 & 8.1 \ \\
500,000 & 34.8 & 41.2 \ & 36.2 & 42.1 \ & 37.9 & 43.2 \ &
39.4 & 43.4 \ & 54.5 & 49.9 \ \\
1,000,000 & 73.2 & 86.8 \ & 76.0 & 88.4 \ & 79.4 & 90.7 \ &
84.3 & 92.8 \ & 144.2 & 119.9 \ \\
5,000,000 & 408.0 & 487.9 \ & 412.6 & 486.9 \ & 441.1 & 505.8 \ &
487.9 & 525.7 \ & 1661.1 & 1056.1 \ \\ \hline
\multicolumn{11}{|c|}{\rule[-2.5mm]{0mm}{7mm}{{\tt Record} (32 bytes) \
random}} \\ \hline
& & & & & & & & & & \\[-2.2ex]
100,000 & 6.7 & 7.8 \ & 7.0 & 8.0 \ & 7.2 & 8.1 \ & 7.4 & 8.0 \ &
10.7 & 10.8 \ \\
500,000 & 42.2 & 49.5 \ & 42.9 & 49.4 \ & 43.5 & 50.4 \ &
46.5 & 50.9 \ & 118.9 & 97.4 \ \\
1,000,000 & 91.9 & 107.7 \ & 92.9 & 107.3 \ & 93.8 & 109.5 \ &
102.0 & 113.3 \ & 324.1 & 247.3 \ \\
5,000,000 & 545.1 & 662.7 \ & 545.6 & 646.1 \ & 545.7 & 642.5 \ &
612.0 & 678.8 \ & 2934.0 & 1983.2 \ \\ \hline
\end{tabular}
\end{center}
}
{\noindent \small Table 3. Results of time measurements on a PC.
Sorting times im milliseconds. For details see text
page~\pageref{reft3}.}
\end{figure}

Measurements were done for data generated as uniformly distributed
(pseudo) random, strictly increasing, strictly decreasing, all equal,
having organ pipe characteristic, and uniformly distributed (pseudo)
random sequences of only two values. Organ pipe characteristic means
that values are strictly increasing up to the middle of the array and
then, as mirror image, strictly decreasing up to the end.

Results are shown in Table 3. Additionally to our favored version that
implements model 5, we present results for models 1, 2, and 3, all of
them with two-way and three-way partitioning, and for two variants of
Heapsort.

The table shows results for integer data (\verb+int+, 4 bytes),
supplemented by those for random sequences of double precision
floating point (\verb+double+, 8 bytes) and record data (32 bytes), the
latter of a type declared as

\begin{verbatim}
   typedef struct {
      int key;
      int data[7];
   } Record;
\end{verbatim}

Already at first glance, one can see that for \verb+int+-arrays,
Quicksort needed sorting times below 0.1 seconds for up to one million
elements, and below 0.5 seconds for five million elements, with the only
exception of the pathological case organ pipe with model 1. A similar
anomaly would have shown up with model 2 (median of three) for both
decreasing values and organ pipe, had we not made the modification
mentioned above.

A further look confirmes that three-way partitioning has its cost
compared to the two-way version except for cases where the number of
distinct values of elements is small compared to \textit{n}.
In our tests these are the cases of all equal values and of random data
of only two different ones, where three-way partitioning exhibits its
power. Second, a greater effort to select the partitioning element
obviously leads to costs higher than could be expected from our results
for numbers of comparisons, which were also confirmed by simulations.
Several reasons may be responsible for this, such as the overhead of
organizing the selection of the partitioning element, and the
non-locality of memory accesses in this connection. It should be kept in
mind, however, that our main objective is to reduce the probability of
bad cases, not to reduce the average sorting time.

\smallskip

We also made time measurements with two Heapsort variants: the original
version (Williams \cite{will64}), labelled ``classic'' in Table 3,
and Bottom-Up Heapsort (Wegener \cite{weg93}), labelled ``BU''.
These methods are known to have worst case time complexity $O(n \log n)$
and, as does Quicksort, they sort in place, but in contrast to Quicksort
do not need additional stack space.

Comparing the sorting times of these two versions, one can see that the
bottom-up variant needs longer for small values of \textit{n}, but---at
least for random and sorted data of distinct values---the time is
increasing more slowly with \textit{n}, so that this version eventually
overtakes the classic variant in speed. In the test cases where the
number of distinct values of elements is small compared to \textit{n},
however, the classic version turns out to be the faster one.

\pagebreak
\clearpage

In all our test cases, the Quicksort version corresponding to model 5
proved to be faster than Bot\-tom-Up Heapsort, though the difference
might not always appear really impressive. For random data, the lead of
Quicksort is increasing with increasing \textit{n}. This effect even
gets larger for data elements of greater size. Further, the ability of
Quicksort, notably the three-way version, to adapt itself to
characteristics of data is much more distinct than that of Heapsort.

\smallskip

If, as usual in theoretical work (and as was our procedure too), the
number of comparisons is taken as measure for the cost of a sorting
algorithm, Wegener's claim that he made in his article \cite{weg93}
(see its title) is correct. As we could see from results of simulation
programs, the numbers of comparisons of elements for random data
in the range \textit{n} = 100,000 to five millions are actually smaller
for Bottom-Up Heapsort than for all our versions of Quicksort.
However, on real-world computers---at least those that are in widespread
use today---the locality of memory access or its absence obviously
influences the performance of programs to an unexpected extent.

\section*{\textnormal{9. \ Conclusion}}

A few years after the fiftieth anniversary of Quicksort, we are able to
present numerical results for the probability of bad cases for the
algorithm. By means of simulations, such probabilities can be determined
only for small numbers of elements. So we computed the frequency
distributions of comparisons by solving the respective recurrence
equations numerically, applying well-known techniques. From these
distributions, probability values for bad cases could be extracted.

For technical reasons, we had to limit our computations to numbers of
elements not greater than \textit{n}~=~500. We believe, however, that
even results in this limited range allow some recognition of the
probabilities we are interested in.

\smallskip

So, how many cases are bad cases? That depends, of course, on how bad
cases are exactly defined, on the number of elements to be sorted, and
on the Quicksort version used. Probability values and their dependencies
are presented in Section 6.4 by means of figures and tables.

\smallskip

The large influence of the selection method on the probability of bad
cases encouraged us to investigate an extension of the median of three
medians selection, which we call ``recursive median of three medians.''
In order to prove that it is applicable in practice, we developed the
implementation that is listed in the appendix. This version not only
reduces probabilities of bad cases to extremely small values and leads
to a worst case complexity below $O(n^2)$, but also helps to prevent
that ``non-random'' data, i.\,e. sequences already possessing some
order, are preferred candidates for bad cases. Time measurements showed
that this version of Quicksort is faster than Heapsort and Bottom-Up
Heapsort for large numbers of elements. For Bottom-Up Heapsort and
random data, this is in contrast to average numbers of comparisons which
we determined by simulations. Theory is important, but is always an
abstraction from reality, so tests are necessary to confirm or disprove
the applicability in practice.

\smallskip

Source files containing the Quicksort version listed in the appendix and
our version of \verb+qsort+ are submitted to the arXiv together with this
article.

\vspace{0.4cm}

\begin{center}
 ACKNOWLEDGEMENTS
\end{center}
{\small
 This work has been done as a private research project of the author
 after retirement from teaching. I am indebted to the Government of
 the Land Schles\-wig-Hol\-stein, Germany, and to our university for
 providing me with the necessary computing resources. Finally, I would
 like to thank our laboratory engineers Rolf Himmighoffen and Maike
 Sieloff for their technical assistance.
}

\newpage

\section*{\textnormal{Appendix: \ Listing of a Quicksort implementation
in C}}

\small

\begin{verbatim}
   /***********************************************************************
   *
   * Name:          quicksort
   *
   * Purpose:       Sorting of an array of element type int into
   *                nondescending order.
   *
   * Prototype:     void quicksort(size_t n, int a[]);
   *
   * Parameters:
   *
   * n:        (in) Number of array elements to be sorted.
   *
   * a:    (in/out) On input, the array to be sorted.
   *                On output, the sorted array.
   *
   *
   * Method:        Sorting is done by nonrecursive Quicksort, using a
   *                stack to store (sub)problems yet to be solved.
   *                For the selection of partitioning elements, a recursive
   *                median of three medians approach is used. Sorting of
   *                small (sub)arrays is done by insertion sort.
   *
   ***********************************************************************/
   #include <stddef.h>
   #include <limits.h>

   #define THREEWAY 1   /* != 0: Use three-way partitioning */

   /*--------------------------------------------------------------------*/
   typedef int Elemtype;
                        /* Comparison operators for Elemtype: */
   #define LT(v1, v2)   ((v1) < (v2))     /* less than */
   #define LE(v1, v2)   ((v1) <= (v2))    /* less than or equal */
   #define EQ(v1, v2)   ((v1) == (v2))    /* equal */
   /*--------------------------------------------------------------------*/

   static size_t medofmed(size_t m, size_t inc, const Elemtype a[]);

   void quicksort(size_t n, Elemtype a[])
   {
   /*--------------------------------------------------------------------*/
      enum {QMIN=5};          /* For recursive median of three medians:
                                   minimum value for (n / sample size)
                                   (must be >= 1) */
      enum {NBASISMAX=15};    /* Maximum (sub)array size for insertion sort
                                   (must be >= 3) */
   /*--------------------------------------------------------------------*/
      enum {FALSE, TRUE};
      enum {STACKSIZE=sizeof(size_t)*CHAR_BIT*2};
      int top, done;
      size_t left, right, nt, i, j, k, ipartel, nc, m, mt;
      Elemtype partval, temp;
      size_t stack[STACKSIZE];

\end{verbatim}

\newpage

\begin{verbatim}
      if (n <= 1)
         top = 0;
      else {
         stack[0] = 0;
         stack[1] = n-1;
         top = 2;
      }
                                 /* loop to deal with stacked intervals
                                      yet to be sorted */
      while (top > 0) {
         right = stack[--top];
         left  = stack[--top];
                                    /* loop to replace tail recursion */
         while (left < right) {
            nt = right-left+1;
            if (nt <= NBASISMAX) {     /* use insertion sort */

               for (i = left+1; i <= right; i++) {
                  if (LT(a[i], a[j=i-1])) {
                     temp = a[i]; a[i] = a[j];
                     while (j > left && LT(temp, a[k=j-1])) {
                        a[j] = a[k]; j = k;
                     }
                     a[j] = temp;
                  }
               }
               left = right;              /* to terminate loop */

            } else {

               if (nt < (QMIN*9)) {    /* median of three */

                  i = left+nt/4; k = left+nt/2; j = k+nt/4;
                  ipartel = (LE(a[i], a[k]) ?
                              (LE(a[k], a[j]) ? k :
                                LT(a[i], a[j]) ? j : i) :
                              (LE(a[j], a[k]) ? k :
                                LT(a[j], a[i]) ? j : i));
               } else {
                                       /* recursive median of medians */
                  nc = nt/(QMIN*9);
                  m = 1;
                  while ((mt = m*3) <= nc)
                     m = mt;
                  m *= 9;                 /* m is sample size */

                  ipartel = medofmed(m, (nt-1)/(m-1), a+left) + left;
               }

               partval = a[ipartel];
   #if (!THREEWAY)                     /* two-way partitioning */
               a[ipartel] = a[right];
               a[right] = partval;
               i = left;
               j = right-1;
               done = FALSE;
               do {
                  while (LT(a[i], partval))
                     i++;
                  while (i < j && LT(partval, a[j]))
                     j--;
                  if (j <= i)
                     done = TRUE;
                  else {
                     temp = a[i]; a[i++] = a[j]; a[j--] = temp;
                  }
               } while (!done);
               a[right] = a[i];
               a[i] = partval;
               j = i;
   #else                               /* three-way partitioning */
               j = left;
               k = right;
               done = FALSE;
               do {
                  while (LT(a[j], partval))
                     j++;
                  while (j < k && LE(partval, a[k]))
                     k--;
                  if (k <= j)
                     done = TRUE;
                  else {
                     temp = a[j]; a[j++] = a[k]; a[k--] = temp;
                  }
               } while (!done);

               k = j;
               i = right;
               done = FALSE;
               do {
                  while (k <= i && EQ(partval, a[k]))
                     k++;
                  while (k <= i && LT(partval, a[i]))
                     i--;
                  if (i <= k)
                     done = TRUE;
                  else {
                     temp = a[k]; a[k++] = a[i]; a[i--] = temp;
                  }
               } while (!done);
   #endif
                                       /* push boundaries of greater
                                            subarray onto stack */
               if (j-left <= right-i) {
                  stack[top++] = i+1;
                  stack[top++] = right;
                  right = (j == 0) ? j : j-1;
               } else {
                  stack[top++] = left;
                  stack[top++] = (j == 0) ? j : j-1;
                  left = i+1;
               }
            }
         }
      }
   }
\end{verbatim}

\newpage

\begin{verbatim}
   /***********************************************************************
   *
   * Name:          medofmed
   *
   * Purpose:       Determination of the recursive median of three medians
   *                from a sample of elements of type int.
   *
   * Prototype:     size_t medofmed(size_t m, size_t inc, const int a[]);
   *
   * Parameters:
   *
   * m:        (in) Number of array elements to be used as sample.
   *                The value must be a power of 3 and >= 9.
   *
   * inc:      (in) Increment of array elements to be used as sample.
   *
   * a:        (in) Array containing elements to be used as sample.
   *
   * Return value:  Index of element containing the recursive median of
   *                three medians.
   *
   ***********************************************************************/
   static size_t medofmed(size_t m, size_t inc, const Elemtype a[])
   {
   #define MEDOF3(i, j, k) \
      (LE(a[i], a[j]) ?    \
         (LE(a[j], a[k]) ? j : LT(a[i], a[k]) ? k : i) : \
         (LE(a[k], a[j]) ? j : LT(a[k], a[i]) ? k : i))

      size_t i0, i1, i2;

      if (m == 9) {
                                       /* recursion basis */
         size_t inc3 = inc*3;
         size_t i = 0, j = inc, k = j+inc;
         i0 = MEDOF3(i, j, k);
         i += inc3; j += inc3; k += inc3;
         i1 = MEDOF3(i, j, k);
         i += inc3; j += inc3; k += inc3;
         i2 = MEDOF3(i, j, k);
      } else {
                                       /* recursive part */
         m /= 3;
         i0 = medofmed(m, inc, a);
         i1 = medofmed(m, inc, a+m*inc) + m*inc;
         i2 = medofmed(m, inc, a+m*inc*2) + m*inc*2;
      }

      return MEDOF3(i0, i1, i2);
   }
\end{verbatim}

\end{document}